\documentclass[journal,comsoc]{IEEEtran}

\pdfoutput=1

\RequirePackage[orthodox]{nag}

\usepackage[T1]{fontenc}
\usepackage{amsmath}
\usepackage[cmintegrals]{newtxmath}
\usepackage[utf8]{inputenc}

\usepackage[caption=false,font=normalsize,labelfont=sf,textfont=sf]{subfig}
\usepackage{url}
\usepackage{booktabs}
\usepackage{acronym} 
\usepackage{array} 
\usepackage{tikz} 

\acrodef{CoE}{certainty-of-event}
\acrodef{PoW}{Proof-of-Work}
\acrodef{LDM}{local dynamic map}
\acrodef{BFT}{Byzantine fault tolerance}
\acrodef{CAM}{cooperative awareness message}
\acrodef{DENM}{decentralized environmental notification message}
\acrodef{V2I}{vehicle-to-infrastructure}
\acrodef{V2V}{vehicle-to-vehicle}
\acrodef{VANET}{vehicular ad-hoc network}
\acrodef{MANET}{mobile ad-hoc network}
\acrodef{WSN}{wireless sensor network}
\acrodef{HSM}{hardware security module}
\acrodef{CAN}{controller area network}
\acrodef{ECU}{electronic control unit}
\acrodef{OBU}{on-board unit}
\acrodef{PKI}{public key infrastructure}
\acrodef{RSU}{road side unit}
\acrodef{ITS-S}{ITS-station}
\acrodef{BSM}{basic safety message}
\acrodef{CA}{certification authority}
\acrodef{ECDSA}{elliptic curve digital signature algorithm}
\acrodef{TTP}{trusted third party}
\acrodef{DSR}{dynamic source routing}
\acrodef{GPS}{global positioning system}
\acrodef{EEBL}{emergency electronic break light}
\acrodef{cITS}{cooperative intelligent transportation system}
\acrodef{CPS}{cyber-physical system}
\acrodef{ABS}{anti-lock braking system}
\acrodef{IDS}{intrusion detection system}
\acrodef{IPS}{intrusion prevention system}
\acrodef{ICS}{industrial control system}
\acrodef{MesAC}{message authentication code}
\acrodef{MedAC}{medium access control}
\acrodef{SPaT}{signal phase and time}
\acrodef{TOPO}{Topology Specification}
\acrodef{PC}{pseudonym certificates}
\acrodef{UMTS}{universal mobile telecommunications system}
\acrodef{LTE}{long-term evolution}
\acrodefplural{cITS}[cITS]{cooperative intelligent transportation systems}
\acrodefplural{CPS}[CPS]{cyber-physical systems}
\acrodefplural{ABS}[ABS]{anti-lock braking systems}
\acrodefplural{IDS}[IDS]{intrusion detection systems}
\acrodefplural{IPS}[IPS]{intrusion prevention systems}
\acrodefplural{ICS}[ICS]{industrial control systems}
\acrodefplural{PKI}[PKIs]{public key infrastructures}

\newcolumntype{C}[1]{>{\centering}p{#1}} 
\newcolumntype{L}[1]{>{\raggedright\arraybackslash}p{#1}}
\newcolumntype{R}[1]{>{\raggedleft\arraybackslash}p{#1}}

\usetikzlibrary{%
	chains,
	calc,
	shapes,
	decorations,
	scopes,
	arrows,
	trees,
	decorations.pathreplacing,
	decorations.pathmorphing,
	fit,
	matrix
}
\newcommand\car{\includegraphics[width=2em]{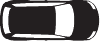}}
\newcommand\carattacker{\includegraphics[width=2em]{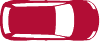}}
\newcommand\carr{\includegraphics[width=2em,angle=180]{img/car}}
\newcommand\laptopattacker{\includegraphics[height=2em]{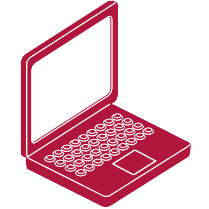}}

\tikzstyle{road-draw}=[draw=black,thick]
\tikzstyle{road-fill}=[black!30]

\newcommand\halfbasicroad{%
\foreach \pos/\coord in {top/3ex, center/0, bottom/-3ex}{
\coordinate (\pos-left) at (0,\coord);
\coordinate (\pos-right) at (0.4\textwidth-\pgflinewidth,\coord);
}

\coordinate (lane1-left) at (1,-1.5ex);
\path (0.4\textwidth-\pgflinewidth,-1.5ex) +(-1,0) coordinate (lane1-right);
\coordinate (lane2-left) at (1,1.5ex);
\path (0.4\textwidth-\pgflinewidth,1.5ex) +(-1,0) coordinate (lane2-right);

\coordinate (lane1) at (lane1-left);
\coordinate (lane2) at (lane2-right);

\fill[road-fill] (top-left) rectangle (bottom-right);
\draw[road-draw] (top-left) -- (top-right);
\draw[road-draw,loosely dashed] (center-left) -- (center-right);
}

\newcommand\basicroad{%
\foreach \pos/\coord in {top/3ex, center/0, bottom/-3ex}{
\coordinate (\pos-left) at (0,\coord);
\coordinate (\pos-right) at (\textwidth-\pgflinewidth,\coord);
}

\coordinate (lane1-left) at (1,-1.5ex);
\path (\textwidth-\pgflinewidth,-1.5ex) +(-1,0) coordinate (lane1-right);
\coordinate (lane2-left) at (1,1.5ex);
\path (\textwidth-\pgflinewidth,1.5ex) +(-1,0) coordinate (lane2-right);

\coordinate (lane1) at (lane1-left);
\coordinate (lane2) at (lane2-right);

\fill[road-fill] (top-left) rectangle (bottom-right);
\draw[road-draw] (top-left) -- (top-right);
\draw[road-draw,loosely dashed] (center-left) -- (center-right);
}

\newcommand\halfroad{%
\halfbasicroad
\draw[road-draw] (bottom-left) -- (bottom-right);
}
\newcommand\road{%
\basicroad
\draw[road-draw] (bottom-left) -- (bottom-right);
}

\newcommand\onramphalfroad{%
\halfbasicroad

\fill[road-fill] (bottom-left) -- ++(3,0) to[out=0,in=180] ++(1,-3ex) -- ++(3,0) coordinate (bend) to[out=0,in=135] ++(1,-3ex) -- ++(45:3ex) to[out=135,in=0] (bend |- bottom-left) -- cycle;

\draw[road-draw] (bottom-left) -- ++(3,0) coordinate (ramp-start) to[out=0,in=180] ++(1,-3ex) -- ++(3,0) coordinate (bend) to[out=0,in=135] ++(1,-3ex) ++(45:3ex) to[out=135,in=0] (bend |- bottom-left) coordinate (ramp-end) -- (bottom-right);

\draw[road-draw,dashed] (ramp-start) -- (ramp-end);
}

\tikzstyle{pencil}=[decorate,
    decoration={random steps,segment length=0.8pt,amplitude=0.3pt}]

\tikzstyle{transmission}=[decorate, decoration={expanding waves, angle=6,
                          segment length=3},
                          very thick,
                          black!10]

\tikzstyle{attack}=[thick,text=black,draw=black,fill=black!05]

\pgfdeclaredecoration{triangle}{start}{
  \state{start}[width=0.99\pgfdecoratedinputsegmentremainingdistance,next state=up from center]
  {\pgfpathlineto{\pgfpointorigin}}
  \state{up from center}[next state=do nothing]
  {
    \pgfpathlineto{\pgfqpoint{\pgfdecoratedinputsegmentremainingdistance}{\pgfdecorationsegmentamplitude}}
    \pgfpathlineto{\pgfqpoint{\pgfdecoratedinputsegmentremainingdistance}{-\pgfdecorationsegmentamplitude}}
    \pgfpathlineto{\pgfpointdecoratedpathfirst}
  }
  \state{do nothing}[width=\pgfdecorationsegmentlength,next state=do nothing]{
    \pgfpathlineto{\pgfpointdecoratedinputsegmentfirst}
    \pgfpathmoveto{\pgfpointdecoratedinputsegmentlast}
  }
}

\tikzset{
    triangle path/.style={decoration={triangle,amplitude=#1}, decorate},
    triangle path/.default=1ex}

\begin{document}

\title{Survey on Misbehavior Detection in Cooperative Intelligent Transportation Systems}

\author{%
Rens W. van der Heijden,
Stefan Dietzel,
Tim Leinm\"{u}ller,
Frank Kargl
\thanks{R. W. van der Heijden and F. Kargl are with the institute of distributed systems, Ulm University, Germany}
\thanks{S. Dietzel is at Humboldt-Universit\"{a}t Berlin, Germany}
\thanks{T. Leinm\"{u}ller is with DENSO Automotive Deutschland GmbH, Germany}
\thanks{Author version of DOI:10.1109/COMST.2018.2873088 -- © 2018 IEEE version at \texttt{https://ieeexplore.ieee.org/document/8477005}}
}

\maketitle

\begin{abstract} 
Cooperative Intelligent Transportation Systems (cITS) are a promising technology to enhance driving safety and efficiency. Vehicles communicate wirelessly with other vehicles and infrastructure, thereby creating a highly dynamic and heterogeneously managed ad-hoc network. It is these network properties that make it a challenging task to protect integrity of the data and guarantee its correctness. A major component is the problem that traditional security mechanisms like PKI-based asymmetric cryptography only exclude outsider attackers that do not possess key material. However, because attackers can be insiders within the network (i.e., possess valid key material), this approach cannot detect all possible attacks. In this survey, we present misbehavior detection mechanisms that can detect such insider attacks based on attacker behavior and information analysis. In contrast to well-known intrusion detection for classical IT systems, these misbehavior detection mechanisms analyze information semantics to detect attacks, which aligns better with highly application-tailored communication protocols foreseen for cITS. In our survey, we provide an extensive introduction to the cITS ecosystem and discuss shortcomings of PKI-based security. We derive and discuss a classification for misbehavior detection mechanisms, provide an in-depth overview of seminal papers on the topic, and highlight open issues and possible future research trends.
\end{abstract}


\section{Introduction}
\label{sec:introduction}

Throughout the field of computer science, securing systems against malicious attackers has become a fundamental requirement for safe, secure, and dependable operation of applications.
Today, professional attacks against systems, which are mounted by large criminal organizations or even governments, are becoming increasingly common~\cite{kargl_insights_2014,Petit2015-Automated}.
At the same time, computer systems are increasingly intertwined with the real world, making them more appealing targets.
The term \acp{CPS} has been coined to encompass systems that are characterized by a large deployment of networked devices equipped with both sensors and actuators~\cite{Mitchell2014}.
They are distinguished from traditional embedded systems, where individual nodes interact with the real world in strongly constrained environments.
In contrast, \acp{CPS} are highly networked, deployed in large regions, and may contain nodes with heterogeneous computational power.
The content transferred in these networks is highly predictable, relating directly to real-world phenomena~\cite{Amin2013-Quest,Mitchell2014}, a fact that enables novel techniques to detect attacks, collectively referred to as \emph{misbehavior detection}.
A prominent example of such a system is a \ac{cITS}, which consists of vehicles, road-side units and back-end systems, and which is the main focus of this survey.
Attack detection in general is an essential second layer of security for networks, especially for widely deployed networked systems in potentially hostile environments, where attackers may have physical access to a subset of the system.
Furthermore, the impact of such attacks is much greater, as they can easily be tailored to cause real-world harm or loss of life.
Therefore, misbehavior detection in both \acp{CPS} and \acp{cITS} is essential for the secure and thus safe operation of these systems.

Cooperative Intelligent Transport Systems are networks designed to provide a variety of benefits~\cite{Uhlemann2018,CSUR-meta-survey}.
These include improved road-safety, greener driving through improved traffic management, support for partially autonomous vehicles, and infotainment services such as traffic information services.
The characterizing communication paradigm of all these applications is that sensors are used to measure real world conditions, which are then communicated over a ubiquitous network.
This network is built up by equipping each vehicle with a wireless interface, creating a dynamic ad-hoc network that can be accessed without further overhead, which is commonly referred to as a \emph{vehicular ad-hoc network (VANET)}.
The VANET can also include infrastructure components, referred to as \acf{RSU}, which are sparsely positioned along the road.
The resulting network that includes sparse infrastructure is referred to as a \emph{vehicular network}.
Vehicles use the VANET to send and receive information, building a \emph{world model} from received messages, which is used for the applications mentioned above.
However, vehicles can also sense local information through a variety of sensors, especially with recent developments in partially autonomous driving.
This information, communicated through vehicle-internal networks, is used for autonomous decision making by the vehicle, either in dedicated driving scenarios or with complete autonomy.
These applications are often supported by additional infrastructure in the back-end through the Internet.
These extensions of a vehicular network, where cooperation between autonomous vehicles and back-end is central, are collectively referred to as \acp{cITS}.

In addition to the benefits discussed above, \acp{cITS} are expected to support the deployment of autonomous driving technologies with cooperation between vehicles.
Both existing applications, as well as envisioned future developments, present a host of new security challenges.
One of these is the integrity and correctness of transmitted information that is exchanged between vehicles that use \ac{cITS} applications.
A wide body of work is aiming to provide this protection, which can broadly categorized into \textbf{proactive} and \textbf{reactive} mechanisms~\cite{Leinmueller2007-Security}.
Proactive security aims to prevent potential attackers from system access, whereas reactive security assumes that malicious activity can be present within the system and needs to be detected and corrected.
Note that in this context, \emph{outsider} refers to an attacker without system access, whereas \emph{insider} refers to a user with system access (i.e., able to transmit legitimate messages), rather than the physical location of the attacker.

More specifically, \textbf{proactive security} refers to any kind of mechanism that enforces a security policy.
This category includes mechanisms such as integrity and authenticity checks (e.g., verifying cryptographic signatures), access control mechanisms and many other systems.
For instance, in \acp{cITS}, the typical approach is to use \acp{PKI} and only issue key material and certificates to vehicles and other authorized entities.
All unauthorized entities are excluded from the system, because their messages do not contain valid signatures.
The state of the art for such \acp{PKI} is discussed in detail in Section \ref{sec:system-security}.
This creates a trusted perimeter that encompasses all authorized entities, requiring potential attackers to possess valid credentials to access the system, reducing the number of attack vectors.
However, if an attacker compromises key material or otherwise manipulates the messages that are transmitted, attacks can still be successful.

Therefore, proactive mechanisms have to be complemented by \textbf{reactive security}.
Reactive security consists of a \emph{detection} and \emph{reaction} step, where attacks that are not prevented by proactive security can be stopped.
Misbehavior detection is firmly positioned in the category of reactive security mechanisms and belongs to the detection step.
In misbehavior detection, the primary goal is to detect such attacks in the \ac{cITS}; this survey aims to provide an overview of existing results in this new context.
In Section \ref{sec:state-of-the-art-attacks}, we discuss the attacks that are still possible in the context of proactive security.
This covers both general cITS and a specific example application, cooperative adaptive cruise control (CACC)~\cite{Amoozadeh-Commag,alipour-fanid_string_2017,jia_survey_2016,vanderHeijden2017-vnc}.
This application is a primary example of new developments in the field of cITS, which has recently gained the interest of the security community.
This example also demonstrates the connection to \acp{CPS} by discussing attacks on both the cyber (i.e., the network) and the physical (i.e., misbehavior by manipulating the process).

Reactive security mechanisms have a long history in traditional networks, where both \acp{IDS} and \acp{IPS} are used.
These are actively researched and many surveys on the topics exist~\cite{Sabahi2008,Lazarevic2005,EstevezTapiador2004,Debar1999}.
A general classification is through the method of detection: either by recognizing known attack patterns (signature-based), detection of anomalous behavior (anomaly-based), or detection of specification deviations (specification-based).
Each of these approaches can be viewed as a classifier of traffic as either benign or malicious.
Misbehavior detection instead classifies messages as correct or incorrect, where correctness refers to whether the messages reflect the real world.
This involves application semantics, which exist beyond the communication that is analyzed (e.g., physical laws and processes in \acp{CPS} and also \acp{cITS}).
An example is the area of routing in ad-hoc networks, where the goal is to detect nodes that exhibit incorrect or malicious forwarding behavior~\cite{Marti2000:Mitigating}.

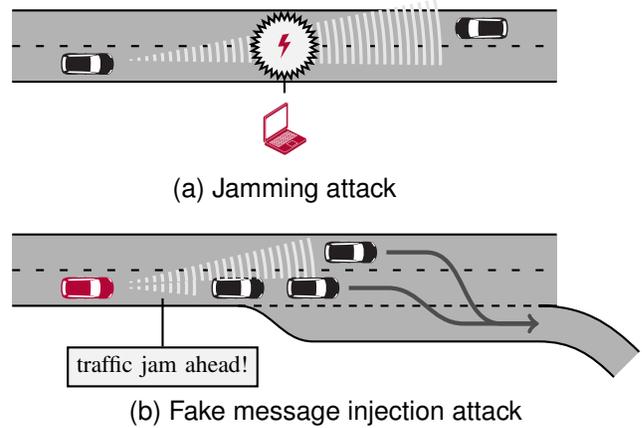
\begin{figure}
    \subfloat[Jamming attack]{%
	\label{fig:jamming-attack}
    \begin{tikzpicture}
        \halfroad
        
        \path (lane1) +(0,0) node (car1) {\car};
        \path (lane2) +(0,0) node (car2) {\carr};
        
        \draw[transmission] (car1) -- node (center) {} (car2);
        
        \node[attack,star,star points=31,star point ratio=0.8,minimum size=2em] (attack) at (center) {\includegraphics[height=1em]{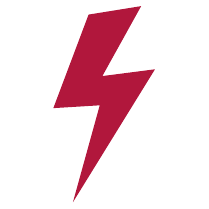}};
        
        \path (center) +(0,-1.1) node (attacker) {\laptopattacker};
        
        \draw[attack] (attacker) edge (attack);
    \end{tikzpicture}
    }
    \hfill
    \subfloat[Fake message injection attack]{%
	\label{fig:fake-message}
    \begin{tikzpicture}
        \onramphalfroad
        
        \node (attacker) at (lane1) {\carattacker};
    
        \path (lane1) +(2,0) node (car1) {\car};
        \path (lane2-left) +(3.5,0) node (car2) {\car};
        \node[right of=car1] (car3) {\car};
        
        \draw[transmission] (attacker) -- node (center) {} (car1);
        \draw[transmission] (attacker) -- (car2);
        
        \path (center) +(0,-1.05)
            node[font=\small,attack] (message) {traffic jam ahead!};
        
        \draw[attack] (center) edge (message);
        
        \draw[->,black!70,ultra thick]
            (car3) -- ++(1,0) to[out=0,in=180] ++(1,-3ex) -- ++(1,0);
        \draw[->,black!70,ultra thick]
            (car2) -- ++(1,0) to[out=0,in=180] ++(1,-6ex) -- ++(0.5,0);
    \end{tikzpicture}
    }
\caption{Two examples for different types of misbehavior.}
\end{figure}

Because of the envisioned large-scale deployment and the direct interaction with the physical world, it is likely that intelligent transportation systems will attract a number of attackers.
Many different types of misbehavior and attacks are conceivable, and we will provide an in-depth discussion and categorization in Section~\ref{sec:misbehavior}.
For now, consider two example attacks to demonstrate the range of possibilities.
Figure~\ref{fig:jamming-attack} shows a jamming attack.
Here, an attacker creates noise on the wireless channel, which hinders the message transfer between two benign vehicles.
To mount this attack, the attacker does not need any particular knowledge about the semantics of the messages that the vehicles exchange.
It is enough for the attacker to know the specification of the wireless communication channel, which is publicly available in standard documents.

Figure~\ref{fig:fake-message} demonstrates a completely different type of attack.
Here, the attacker uses knowledge of the protocol semantics to create a message that informs the vehicles ahead of an (non-existent) traffic jam.
Receiving vehicles will use this message for routing decisions and might take an alternate route.
As a result, there might be less traffic on the original road, which is a direct benefit for the attacker.
The main harm caused in this case is that vehicles might slow down when approaching the non-existent traffic jam location, maybe creating a real traffic jam.
In addition, the traffic participants might get confused in case they notice that there is no traffic jam on the original road. As a result the user acceptance of the system is lowered.

If we consider safety applications or automated driving~\cite{Petit2015-Automated}, it is conceivable that attacks on intelligent transportation systems can even lead to accidents, possibly threatening the lives of passengers.
For instance, given a set of vehicles that are using a platoon with reduced relative safety distance, an attacker could inject a fake message that indicates one of these vehicles is breaking, even though this is not the case.
In such a situation, vehicles will immediately respond and in the worst case actually cause collisions to avoid the non-existent breaking behavior claimed by the attacker.
This type of attack has been studied for its effectiveness in recent works~\cite{Amoozadeh-Commag}, but suitable countermeasures require the ability to detect false information~\cite{vanderHeijden2017-vnc}.
This is exactly what misbehavior detection is designed to do.

\section{Related Surveys \& Contribution}\label{sec:related-surveys}

There is a large variety of surveys~\cite{CSUR-meta-survey} that discuss various aspects of \ac{cITS} security, including authentication schemes, attack techniques, routing security, trust management, and revocation.
Compared to these works, we are more focused on misbehavior detection as it relates to real-world processes, and we aim to provide a detailed categorization of existing work.
We also place all of the works into the same \ac{cITS} context, which reveals some vulnerabilities related to assumptions on the attacks that no longer hold.
This categorization thus provides a much deeper insight into the developments of the field over the past two decades, compared to existing work.

There are recent surveys on the security of areas of which \ac{cITS} can be considered a subset, such as trust in \acp{MANET}~\cite{Movahedi2016}, opportunistic networking~\cite{Wu2015}, and \acp{CPS}~\cite{Giraldo2017,Mitchell2014}.
Although these clearly also address issues related to \ac{cITS}, our survey has limited the scope to be able to address the topic in more detail, considering the area of misbehavior detection more specifically.
Other interesting surveys that have synergies with this one include a study on attacks and countermeasures for GNSS spoofing~\cite{Schmidt2016}, a system that is essential for safe operation of \acp{cITS}, and a study on social network cooperation mechanisms in \acp{cITS}~\cite{Hua2017}.
However, none of these surveys directly address security in \acp{cITS} as a primary focus; the unique requirements call for different security mechanisms.

Recently, a number of authors have also written surveys addressing security in \acp{cITS} specifically.
Azees~et~al.~\cite{Azees2016} primarily focus on VANETs as a special case of \acp{MANET}, a view that was common in the past, but which does not include in its' scope the most safety-critical classes of attacks.
Instead, the authors are discussing at routing-oriented attacks and defence mechanisms, while we focus on examining misbehavior detection with respect to \acp{cITS} applications.
Hasrouny~et~al.~\cite{Hasrouny2017} and Sakiz~\&~Sen~\cite{Sakiz2017} both address the \acp{cITS} proper, but these surveys both place a much different focus.

The work by Hasrouny~et~al.~\cite{Hasrouny2017} can be considered a broad introduction to the field, more akin to a tutorial: the authors study the different attack techniques in a lot of detail.
They continue with an extensive description of standardization efforts with respect to security.
The authors finally describe potential countermeasures for the discussed attacks.
However, due to the very broad scope of their work, they cannot go into sufficient detail on detection specifically, and instead provide scattered solutions for individual attacks.
Our survey instead aims at a much deeper understanding of the methods used for misbehavior detection, classifying by different scheme properties.
In particular, we identify patterns in how the literature tackles the topic of detection, and also classify the privacy and generalizability of each scheme.

Sakiz~\&~Sen~\cite{Sakiz2017} also provide a survey of detection mechanisms for \acp{cITS}: however, their scope of detection mechanisms includes the largely parallel topic of routing security.
This is illustrated by the emphasis on denial of service, replay and network layer attacks.
For this reason, they are not able to go into as much detail as we are able to on safety-critical attack types, such as data injection attacks.
Unlike their survey, we provide a qualitative discussion of the detection mechanisms, reflecting on their generalizability and what the weaknesses of each approach are.
Our classification focusses on shared characteristics, which helps technical readers to identify which schemes are comparable, and it helps researchers identify parallels between different schemes, as well as gaps for future work.

In this article, focus on the specific area of misbehavior detection, providing a comprehensive survey of different misbehavior detection mechanisms for \ac{cITS}.
We also consider the implications of this research for other \ac{CPS} domains.
Due to the challenging network topology and application-specific networking paradigms applied in \ac{cITS}, we believe that misbehavior detection mechanisms for these systems offer interesting properties that can be transferred to other use cases.
In particular, the issue of data injection attacks is a challenge that is also observed in other \ac{CPS}, while routing attacks such as those covered by previous surveys are network-specific.
Our main contributions can be summarized as follows:
\begin{itemize}[noitemsep,nolistsep]
\item We start with a comprehensive overview of the state of the art and current standardization efforts, including overall \ac{cITS} security in Section~\ref{sec:system_model}.
\item We define misbehavior, an attacker model, and a novel taxonomy for misbehavior detection in Section~\ref{sec:misbehavior}.
\item We provide an extensive overview of seminal misbehavior detection works in Section~\ref{sec:state_of_the_art}, using the taxonomy as a guideline\footnote{An earlier version of this paper was pre-published on Arxiv~\cite{vanderHeijden2016-Survey}}. We also provide a more fine-grained classification by how detection is performed, which helps readers identify relevant classes of mechanisms that have orthogonal functionality.
\item We discuss solved and open challenges (Section~\ref{sec:solved_and_open_challenges}) and point out commonalities and differences of \ac{cITS} misbehavior detection and misbehavior detection in other domains (Section~\ref{sub:applications_beyond_cits}).
\end{itemize}

\section{System Model}\label{sec:system_model}

\subsection{cITS overview}\label{sub:its_communication_patterns}

The \ac{cITS} system consists primarily of \ac{V2V} and \ac{V2I} communication, which is proposed to be based on the IEEE 802.11p amendment, which specifies MAC and PHY layers for short-range communication between vehicles.
Communication is performed at 5.9 GHz with a communication range that is typically between 100 and 500 meters, and which is highly dependent on the scenario and buildings in the vicinity.
IEEE 802.11p has gained world-wide adoption as a basis on top of which standardization agencies are building \ac{cITS} applications.
For example, the European telecommunications standardization organization, ETSI, has a family of standards for \acp{cITS}, referred to as ITS-G5, which provides communication primitives. 
On top of these, safety and entertainment applications are being developed.
Similar initiatives exist in the US in the IEEE 1609 family of standards, in conjunction with the Society of Automotive Engineers (SAE).

In contrast to routing in classical IT systems such as the Internet, multicast and broadcast are the predominant networking patterns in \ac{cITS}~\cite{Schoch2008}.
These patterns are much more suitable for the envisioned applications, which vary from lane change warnings to cooperative adaptive cruise control (CACC) and city-scale traffic flow optimization.
CACC is an application that has gained significant attention in recent years, and is essentially an extension of adaptive cruise control (ACC), where the safety distance between vehicles is reduced.
In CACC, vehicles periodically exchange position information to form a very tight formation that would normally be susceptible to collisions~\cite{jia_survey_2016}.
This is is a form of partially autonomous driving that goes beyond the potential benefits from sensor-equipped vehicles.
The data exchanged between the network participants, vehicles and infrastructure alike, share the property that they are relevant to all receivers.
Therefore, addressing packets to specific vehicles does not make sense; instead, addressing refers to the local neighborhood (1-hop broadcast), specific regions (geocast) and infrastructure.
This style of addressing allows the network to exploit the simple fact that wireless networks are a broadcast medium by nature.

These communication patterns are supported by two message types in the European model: the periodic \acf{CAM}~\cite{etsi:CAM} and the event-driven \acf{DENM}~\cite{etsi:DENM}.
In the US, standardization foresees a two-part \acf{BSM}~\cite{BSM,SAE:J2735} that roughly corresponds to the \ac{CAM} and \ac{DENM}.
These message types make up the bulk of communication between vehicles; in Europe, additional message types are currently under standardization for specific applications, such as the \acf{SPaT} and \acf{TOPO} message types.
Both the \ac{CAM} and the first BSM part consist of core elements including position, heading, speed, steering wheel angle and vehicle size.
Supplemental information can be added, such as vehicle role and status of vehicle lights.
CACC applications can technically work with just CAM/BSM.1 information, but most proposals include acceleration information into beacon messages for better performance.

The second BSM part is only transmitted when a specific event occurs, similar to \acp{DENM}, but is packaged within the next periodic BSM instead of being transmitted separately.
This second BSM part is also single-hop; unlike the European model, which specifically foresees \acp{DENM} as multi-hop messages, events are transferred through 1-hop BSMs only.
DENMs are event triggered messages, which are designed to warn about specific events, such as traffic jams, emergency breaking, an approaching emergency vehicle, or road construction.
\acp{DENM} are usually relevant for specific geographical area, and can be forwarded there over multiple hops.
A geobroadcast protocol is specified~\cite{etsi:ts102636-4-1}, which first forwards messages to the designated target region and then broadcasts the message within that region.

Although \acs{V2V} communication using \acp{CAM} and \acp{DENM} is able to support a large number of use cases, some use cases require \ac{V2I} communication.
Infrastructure can help to increase communication range during the initial deployment phase, especially in critical areas such as intersections.
Another important feature of infrastructure is that it enables communication with back-end infrastructure and the general Internet.
This is useful for non-safety application classes, such as infotainment services, over-the-air software updates and security credential updates.
\acp{RSU} can also be used to enable specific applications, such as traffic and fuel management through \ac{SPaT} messages\footnote{These messages inform incoming vehicles when the traffic light will turn green, allowing more efficient fuel management.}.
However, large-scale deployment of roadside units is considered unlikely due to the estimated deployment and operation costs of 3,000--5,000 US dollars per \ac{RSU}~\cite{Banerjee2008-rsu-cost}.

Complementary to \acp{RSU}, cellular communication can be used to provide back-end and Internet connectivity, making use of the fact that broadband cellular communication services like \ac{LTE} have already been deployed on a large scale.
However, the necessary developments to support a large fleet of vehicles in addition to regular cell phones have not yet been made, despite recent strides in this area~\cite{Uhlemann2018}.
One of the major open question is related to the business model; thus many assume that cellular support will likely remain restricted to more expensive vehicles, at least as long as cellular usage still incurs considerable fees.
New proposals from the area of 5G suggest that an ad-hoc-like mode dedicated for inter-vehicle communication will also be available in the form of LTE-V, which is inspired by Device-to-Device communication (D2D)~\cite{5G-V2V}.
This approach has significant backing from industry, but lacks fundamental privacy guarantees in the same way that standard cellular communication does.
Independent of C-V2X, some have proposed the use of a heterogeneous network in recent years, using both IEEE 802.11p-based and cellular communication~\cite{hetnets}.
This approach can take advantage of both networks, but also presents the challenge of efficiently managing the available channels in a decentralized way~\cite{LTE-DSRC-traffic-steering}.
Both heterogeneous networks and pure cellular proposals are still in research phases, and will likely face many of the same security challenges as IEEE 802.11p.
Although our survey focuses primarily on the IEEE 802.11p setting, we expect the results of many schemes to transfer to C-V2X with limited modifications.

Beyond the communication model under standardization, some authors propose that clustering might be a viable approach to organize the communication.
Current standards do not foresee this approach because it has significant disadvantages in terms of communication overhead to create and maintain clusters~\cite{Willke2009-protocols}.
These clusters are typically created and maintained in an ad-hoc fashion, to facilitate deployment, which requires vehicles to agree on one or more cluster heads that manage a cluster.
The disadvantage of this approach is that it requires a fundamental inequality between different vehicles to take advantage of clustering, i.e., the cluster heads must have an authoritative status.
Because this network model security challenges that are widely different from the standard \ac{cITS} communication models, we do not address them in detail in this survey.

\subsection{cITS Security}\label{sec:system-security}

In order to mitigate possible attack vectors, security features are actively considered by both academia and industry.
Standardization agencies have developed an initial standard for practical implementation, which serves as the basis for a lot of work in \ac{cITS} security.
Unlike traditional networks, though, confidentiality is not considered a major requirement, because most applications inherently rely on the data exchanged between vehicles.
The main focus is instead to ensure integrity and authenticity, for which standardization employs proactive security mechanisms.
The basic approach is to set up a \acp{PKI} and provide each vehicle with an asymmetric key pair and a certificate.
The certificate contains the public key alongside a number of \ac{cITS} specific attributes (e.g., vehicle type, vehicle dimensions, license plate number) and is signed by the key issuing authority~\cite{Raya2007-Securing,Kargl2008,Papadimitratos2008}.
This serves as a long-term identifier of a vehicle that confirms it as a valid participant in the \ac{cITS}.
We first discuss the basic application of these certificates for cITS security, before discussing proposals for credential management in detail in Section \ref{sub:its-privacy}.

The ETSI TS 103 097 standard~\cite{etsi:ts103097} in Europe and the IEEE 1609.2 standard~\cite{ieee:1609.2} in the US specify how certificates should be used to achieve integrity and authenticity.
The basic idea is that each outgoing message is signed using the sender's secret key.
The signature and certificate are both attached to the message to enable broadcast authentication, i.e., each receiver can check message authenticity without further message exchanges.
For efficiency reasons, some authors have proposed certificates could be omitted periodically to save bandwidth~\cite{Feiri2012-Congestion}, but this introduces additional delays beyond those already present due to time needed for verification.
In both European and American standards, the cryptographic algorithm is the \acf{ECDSA}, using the p-256 curve standardized by NIST~\cite{nist:fips.186-4}.
This signature and certificate validation process provides sender authenticity and message integrity, protecting against attackers that transmit messages using commodity hardware without key material.

While signature and certificates effectively thwart most attacks using commodity hardware without key material, they do not provide any guarantees on message correctness.
Because of the expected wide deployment of \ac{cITS} and the long lifetime of vehicles, it is likely that attackers either physically own a vehicle or are otherwise able to extract key material from (old) communication units.
If key material is stored on a regular storage medium (e.g., hard disk or flash memory), attackers with physical access to a car could easily extract it, transfer it to other devices, and create arbitrary messages with valid signatures.
Some propose to use trusted hardware, usually in the form of a hardware security module, to protect key material~\cite{Raya2007-Securing,Kargl2008}.
Here, the idea is to store (secret) key material within a tamper resistant component that is protected against outside access.
If a message needs to be signed, it can be forwarded to the trusted device and the signature is returned.
Thus, the key material never leaves the trusted hardware, making it much harder for an attacker to extract secret keys and subsequently sign arbitrary messages.
As in normal \acp{PKI}, revocation can be employed to mitigate this issue, but this raises new questions on how key material abuse can be detected in existing vehicles.

Note that if key material is kept in trusted hardware, the application code is still untrusted: it is likely that attackers are able to manipulate the software in order to create arbitrary messages.
These messages will then be signed, because the signing functionality running on trusted hardware has no way to tell whether messages have been manipulated~\cite{Raya2006-Securing}.
The alternative would be to build an architecture in which all hardware is trusted, but this also has a number of disadvantages.
First, running all software on trusted hardware will greatly increase cost, because more powerful trusted hardware is needed.
In addition, if all applications need to be manually certified and deployed on trusted hardware within the vehicle, over-the-air updates are more complicated and the deployment process of new software is slowed down.
Even if attackers are not able to modify the software, they may be able to modify sensor readings, which lead to modified messages as a result.
To alter sensor readings, attackers can either inject false readings in the CAN bus or directly modify sensor hardware.
Therefore, it is likely that software running on the vehicle can be controlled by a capable attacker, despite trusted hardware.
This requires us to consider correctly signed messages with invalid contents circulating in the network. 

\subsection{Certificate Management \& Privacy}\label{sub:its-privacy}

If vehicles attach their long term certificates to all outgoing messages, they can be tracked using a trace of received messages with attached location information.
For this reason, long-term certificates can be replaced by short-term identifiers, referred to as pseudonyms.
These pseudonyms make it harder to collect location traces of specific vehicles~\cite{Kargl2008,Papadimitratos2008,Petit2014-Survey}, while still providing the same authenticity guarantees.
While pseudonyms enhance privacy, their parallel use enables potential attacks, often referred to as Sybil attacks~\cite{Douceur2002-Sybil}.
This is one of the classes of attacks that we will discuss extensively in this survey (in section \ref{sec:attacker-model}), as it is one of the key challenges that misbehavior detection helps address.
However, in this section we first present a discussion of some proposals for certificate management.

A standard \ac{PKI} only foresees the model as discussed above: each entity has a valid certificate that identifies the entity, and authenticity entails identification.
For privacy reasons, many proposals have been made to issue a set of pseudonyms, which is typically linkable under certain conditions to the long-term identifier of the entity~\cite{Petit2014-Survey}.
The creation, distribution, management and revocation of these pseudonyms present a trade-off between privacy, security and performance.
Two major proposals are now in the initial stages of deployment, which analogous to standardization of communication can also be seen as a European and a US proposal.
In the US proposal, Brecht~et~al.~\cite{Brecht2018-SCMS} propose the use of \emph{butterfly keys} to provide a cryptographic link between different certificates.
In this model, pseudonyms are issued in bulk, revocation is done by revealing the butterfly key, and a misbehavior authority interacts with the pseudonym issuing entities to detect attacks in the back-end.
Recent European proposals, such as that suggested by~Khodaei~et~al.~\cite{Khodaei2018}, take an on-demand approach to pseudonym issuing, and claim to have more protection against Sybil attacks.
Such protection can be provided by limiting the validity of pseudonyms, which is a trade-off against performance in the US model (where many pseudonyms would be wasted due to batch issuing).

In general, these pseudonym issuance systems provide a partial framework that in turn influences the suitability of certain detection schemes.
At the time of writing, it is still not clear which exact scheme will be implemented world-wide, or whether disparate systems will be used in different parts of the world.
However, the constraints posed by the credential management system impacts potential detection mechanisms, primarily through how easily Sybil attacks can be executed.
This aspect of pseudonym management is addressed in our survey through a classification, as discussed in Section \ref{sec:pseudonyms}.

\subsection{Attacks on cITS}\label{sec:state-of-the-art-attacks}

Here we briefly review a number of proposed attacks on the cITS, particularly those that can still be performed within the scope of existing security mechanisms as described in the previous subsections.
These are divided along the lines of the cyber and physical in a cyber-physical system; in this survey we primarily focus on the cyber component.

\subsubsection{Cyber attacks}

The space of possible attacks on computer systems is vast; in this survey, we concentrate on attacks that are not easily avoidable through security mechanisms that are already implemented.
As there are many survey articles discussing attacks~\cite{CSUR-meta-survey,Hasrouny2017,Sakiz2017}, we here provide a review of the important types and refer interested readers to other surveys for a detailed review.
The attacks we discuss here are jamming attacks, data injection attacks, replay attacks, routing attacks and Sybil attacks.

The jamming attack was already briefly introduced in Figure \ref{fig:jamming-attack}.
The attacker disrupts communication from or to specific nodes by transmitting a constant or targeted burst communication to disrupt communication between nodes.
This essentially means that the attacker can decide which messages will arrive, and which do not.
This attack is commonly discussed in CACC and can be executed even by low-power devices such as UAVs~\cite{alipour-fanid_string_2017}.

Replay attacks, on the other hand, refer to an attacker re-transmitting received messages.
Standard proactive mechanisms provide some protection against replay attacks by including a time stamp in every message.
In cITS, a relatively reliable source of timing is available in the form of a global navigation satellite system (GNSS), which also provides high-accuracy time synchronization~\cite{Schmidt2016}.
However, in multi-hop communication it is conceivable that these attacks can affect throughput.

Replay attacks in multi-hop communication are closely linked to routing misbehavior, a well-studied class of attacks that aims to disrupt network performance.
This category of attacks originates from MANETs, and both attacks and defense mechanisms have since been transferred to VANETs and other types of ad-hoc networks, as discussed in Section \ref{sec:related-surveys}.
Basically, the attacker deviates from the routing protocol to cause messages to be lost or be routed through a specific node.
Unlike most attacks we study, routing attacks do not directly affect safety, and thus we consider these attacks to be of limited importance.
Note also that cITS includes infrastructure; if available, using infrastructure by default for long-distance communication significantly reduces the impact of routing misbehavior.

Another important class of attacks is that of data injection; in these kinds of attacks, an attacker claims information that contradicts with real-world information.
The purpose of such an attack can be disrupting road traffic, or even triggering a collision.
For example, attackers may falsely claim that a road is blocked to have the road for themselves, or they may claim a vehicle is directly in front of a victim, causing the victim to trigger the breaks.
As suggested by these examples, attacks have widely different goals, with varying time-sensitivity.
Because of the variety of attacks, many different mechanisms to detect them have been proposed in the literature, and a large chunk of misbehavior detection is directly related to these attacks.
Authors have shown~\cite{Amoozadeh-Commag,vanderHeijden2017-vnc} that CACC is particularly vulnerable to these attacks, and being able to detect them is essential for the design of secure control algorithms.

Sybil attacks~\cite{Douceur2002-Sybil} are a specific type of attack associated with the pseudonym discussion above.
In the simplest case, an attacker here uses multiple identities to achieve a specific goal.
In cITS, this goal is often related to the application, similar to data injection attacks: an attacker may use multiple pseudonyms to make it appear as though the road is full, even though it is not.
Closely related to this are various types of attacks on specific detection mechanisms, where the attacker uses the identities to boost the reputation of specific vehicles, or conversely reduce the reputation of legitimate vehicles (\emph{bad mouthing attacks}).
A detailed discussion of reputation systems and attacks on them was provided by Koutrouli~et~al.~\cite{koutrouli_reputation_2015}.

\subsubsection{Physical attacks}

Apart from the cyber component, there are also many attacks possible on the physical processes surrounding the vehicle itself.
Recent work on the security of controller area networks (CANs)~\cite{koscher_experimental_2010,kleberger_security_2011}, which are the primary type of bus used in current vehicles, has shown that they are vulnerable to attacks.
So far, this work has primarily focused on attacking a vehicle from the outside; however, an attacker attempting to attack the cITS could similarly exploit their physical access to the bus.
Such attacks allow an attacker to force the on board unit to transmit arbitrary messages that conform to the network specification without needing to compromise the hardware security module.

Some authors have also suggested that physical attacks on specific vehicles may be possible.
This works by directing the attack towards the sensors themselves~\cite{Petit2015-Automated}.
A simple example of such an attack would be taking a laser pointer and disrupting the cameras of a nearby vehicle.
In this survey we primarily focus on attacks in the cyber space; however, we feel it is important to also consider that physical sensors are not necessarily immune to attacks.

\section{Misbehavior in cITS}
\label{sec:misbehavior}

In this section, we will go into more detail concerning misbehavior in \acp{cITS}.
In particular, we discuss the attacker model, different types of realistic attackers and a taxonomy of misbehavior detection.
As detection can be performed on different scopes, we also briefly address this topic and the closely related topics of misbehavior reporting and revocation.
Finally, we will go into the topic of pseudonyms, in order to investigate the compatibility of the misbehavior detection mechanisms with this privacy-enhancing mechanism.
We will use each of these aspects in our classification of the literature in the next section.

\subsection{Definition and Attacker Model}
In this subsection, we will treat the definition of misbehavior and the attacker model associated with it.
We distinguish these two concepts because misbehavior gives a strong intuitive understanding of the ultimate goal of our work.
The attacker model aims to provide a more formal version of this intuition, based on the state of the art.

\subsubsection{Defining Misbehavior}
In literature, there are many different definitions of misbehavior~\cite{Raya2006-Securing,Raya2007-Securing,Raya2007-Eviction,Zhuo2009,Rivas2011}, often implicitly defined by the attacker model rather than explicitly being stated alongside it.
Misbehavior is a broad term; its origins are somewhat unclear, but it is commonly used for ad-hoc networks when discussing specific attacks that are executed by the participating nodes, as opposed to intrusions or attacks.
Our definition of the term covers not only attackers and malicious participants, but also faulty nodes.
\emph{Misbehaving nodes} are thus any node that transmits erroneous data that it should not transmit when the hard- and software are behaving as expected.
We argue that a misbehaving node is the type of node we should detect.
However, the literature often distinguishes~\cite{Leinmueller2007-Security,Lo2007,Kargl2008,Lin2008-Security} between \emph{faulty node} and \emph{malicious node}, which are defined as follows:

\emph{Faulty nodes} are those participants in the network that produce incorrect or inaccurate data without malicious intent.
Faulty nodes are usually related to faults in sensors, either because the sensor was damaged or because of errors caused by variance in the sensor.
Examples include a heat sensor that is malfunctioning, causing the node to transmit the current temperature to be -50 degrees Celcius, and the error caused by a reading produced by a GPS device, which can cause vehicles to transmit an erroneous position.

\emph{Malicious nodes} or attacker nodes are those nodes that are transmitting erroneous messages with malicious intent.
Such messages may also be referred to as deceptive messages, and the attackers as deceptive attacks.
These nodes are our main target for detection, as they represent a direct threat to the integrity of the data exchanged in the network.
These nodes can actively attempt to avoid detection, or subvert other nodes in the network to transmit their erroneous messages.
This definition includes denial of service attacks and exclusion of messages that should be sent (e.g., attacks on routing).
The goal, source and means of these nodes may vary greatly, as discussed below.

Note that these definitions are not consistently used throughout the literature; in our work, we use misbehavior detection to refer to the detection of both faulty and malicious nodes.
Unlike traditional intrusion detection, misbehavior detection attempts to detect incorrect packets, rather than detecting malicious packets.
There are many good reasons to avoid a focus on content in a generic networking setting: unlike in \acp{cITS}, such networks see a lot of encrypted traffic and a large variety of application layer protocols.
In \acp{cITS}, we have relatively few protocols, and the public nature of transmitted data implies that content is rarely encrypted.
In addition, \acp{cITS} lack a clear system boundary, which is essential for effective \ac{IDS} deployment.
Misbehavior detection can take advantage of the public nature of data, and does not suffer from the lack of a system boundary, making it a potentially more suitable approach.
This type of detection may not be feasible for normal networks, but the characteristics shared between \acp{cITS} and \acfp{CPS} suggest some schemes could be used in both systems.

\subsubsection{Attacker Model}\label{sec:attacker-model}
We now review several attacker models that are used in the literature.
The secondary purpose of this section is to discuss the challenges involved with the development of a good and universal attacker model for \acp{cITS}.
Unlike most network scenarios, there is no universally accepted attacker model that is consistently used for \acp{cITS}; here we provide a brief overview of the most common assumptions.

The attacker model from~\cite{Raya2007-Securing}, one of the seminal works on security in \acp{cITS}, presents the following four classification dimensions for attackers, and a variety of basic and sophisticated attacks.
We already discussed the distinction between \emph{insider} and \emph{outsider}, i.e., whether or not the attacker possesses valid credentials.
The motivation of the attacker is classified as either \emph{rational}, where the attacker has a direct benefit from his attack, and \emph{malicious}, where an attacker only seeks to disrupt or cause harm.
The attacker may be \emph{active} or \emph{passive}, which considers whether the attacker can transmit messages or signals, or whether the attacker only eavesdrops on the channel.
For this survey, we seek to detect active attackers only, as the very goal of misbehavior detection is to determine whether certain messages or signals constitute undesirable behavior; eavesdropping is not misbehavior, in the sense that it cannot be detected.
The final classification dimension is the scope of the attack, which may be \emph{local} or \emph{extended}.
This distinction does not consider the amount of attacker-controlled nodes, but rather their \emph{distribution} over the network.
Local refers to one or more attackers distributed in a small region, such as a highway section between two intersections or a few neighboring intersections in an urban setting.
The extended scope provides for a number of attacker-controlled nodes across a larger region.

Most attackers in this classification are instances of the standard Dolev-Yao attacker model.
However, the insider attackers that behave according to the protocol, but send false information are not covered in the attacker model.
Similarly, as discussed by the authors~\cite{Raya2007-Securing}, Sybil attacks play a significant role.
In those attacks, attackers obtain multiple identities (multiple certified key-pairs) to circumvent reputation mechanisms.
Many authors of detection mechanisms implicitly or explicitly limit the potential for Sybil attacks, meaning the attacker has possession of a small set (usually 1-3) of valid pseudonyms for each vehicle they control.
Because this is usually done through the \ac{PKI}, there is a potential impact on privacy, which many authors consider acceptable to guarantee safe operation of the \ac{cITS}.

One concern is that many authors~\cite{Raya2007-Securing} consider \acp{RSU} to be fully trusted.
We remark that although this may be the case for connections leading outward through the \acp{RSU}, full trust cannot be assumed for the devices themselves.
As the name implies, \acp{RSU} are deployed on the side of the road, which makes them susceptible to physical attacks like sensor tampering and differential power analysis, just like regular vehicles.
Because \acp{RSU} are potentially more authoritative, they are potentially valuable targets if too much trust is placed on them.

On the other hand, the full Dolev-Yao model, with the extensions and limitations described above, is too strong to provide a meaningful level of security.
Particularly, if the attacker is allowed to arbitrarily re-organize the packets received by each individual node in the network (as in the Dolev-Yao model), they can selectively deliver only those messages that confirm the view presented by the attacker.
In reality, the attacker is also subject to the limitations of physics.
For example, an attacker cannot arbitrarily re-arrange the reception of messages for each receiver in a wireless network.
For this reason, the literature specifies an important limitation compared to the Dolev-Yao model: an attacker must be a node in the network, restricted by the same physical properties that restrict a normal node.
Although their transmission range may be increased through power control~\cite{yao_multi-channel_2018}, attackers are considered to have a limited physical presence~\cite{Papadimitratos2008}.
In addition, it is commonly assumed that attackers are constrained to physical limitations such as the speed of light.
Similarly, an attacker cannot in general receive and jam a message at the same time without the risk that some other receiver also successfully receives the message.
Another issue with the Dolev-Yao model is that it is not designed for the analysis of attacks on application semantics, i.e., message correctness is not considered, which is the primary focus of misbehavior detection.

In the literature, some authors have attempted to formalize these attacks and the effects on specific mechanisms using game theory~\cite{Bilogrevic2010,Liu2010}.
These formalizations often make an informal honest majority assumption, which assumes that a majority of the nodes (or sometimes keys or messages) is honest.
The assumption is sometimes explicitly specified as being about a local neighborhood or the entire network.
A local honest majority assumption is then the assumption that more than half of the nodes in the direct communication range of a vehicle is honest.
This assumption is the most significant limitation of some classes of detection mechanisms, which we will discuss at length in this survey.
Apart from this, we will rely on terminology introduced by earlier work~\cite{Raya2007-Securing,Bissmeyer2011-Simulation,Papadimitratos2008,Leinmueller2008-Modeling}.

\subsection{Scope of Detection}\label{sec:scope}

\begin{figure*}
  \centering
\begin{tikzpicture}[node distance=1em]
\tikzstyle{box}=[draw=black,fill=black!20,minimum width=8em,minimum height=3ex];
\tikzstyle{confidence}=[box,fill=black!20];
\tikzstyle{consequence}=[box,fill=black!05];

\begin{scope}[start chain=going below]
    \node[box,on chain] (sensors) {Sensors};
    \node[box,on chain] (lwm) {World model};
    \begin{scope}[start branch=detect going right]
        \node[confidence,on chain] (ldetection) {Detection};
        \begin{scope}[start branch=report going below]
            \path node[confidence,on chain] (lreporting) {Reporting};
            \begin{scope}[start branch=revoke going right]
                \node[consequence,on chain] (lrevocation) {Revocation};
            \end{scope}
        \end{scope}
    \end{scope}
    \node[box,on chain] (ldissemination) {Dissemination};
    
    \node[on chain] {};

    \node[box,on chain] (nwm) {World model};
    \begin{scope}[start branch=detect going right]
        \node[confidence,on chain] (ndetection) {Detection};
        \begin{scope}[start branch=report going below]
            \path node[confidence,on chain] (nreporting) {Reporting};
            \begin{scope}[start branch=revoke going right]
                \node[consequence,on chain] (nrevocation) {Revocation};
            \end{scope}
        \end{scope}
    \end{scope}
    \node[box,on chain] (ndissemination) {Dissemination};

    \node[on chain] {};

    \node[box,on chain] (gwm) {World model};
    \begin{scope}[start branch=detect going right]
        \node[confidence,on chain] (gdetection) {Detection};
        \node[consequence,on chain] (grevocation) {Revocation};
    \end{scope}
\end{scope}

\tikzstyle{every edge}=[draw,thick];

\draw[->] (sensors) edge (lwm)
          (lwm) edge (ldissemination);

\path[draw,->] (ldissemination) edge (nwm);
\path[draw,->] (nwm) edge (ndissemination);
\path[draw,->] (ndissemination) edge (gwm);

\path[<->] (lwm) edge (ldetection);
\path[<->] (nwm) edge (ndetection);
\path[<->] (gwm) edge (gdetection);

\path[->] (ldetection) edge (lrevocation);
\path[->] (ldetection) edge (lreporting)
          (lreporting) edge (nwm)
          (ndetection) edge (nrevocation);
\path[->] (ndetection) edge (nreporting)
          (nreporting) edge (gwm)
          (gdetection) edge (grevocation);


\tikzstyle{box}=[thick,draw=black!80!black];

\path (sensors.north west) +(-3pt,3pt) coordinate (local-tl);
\path (ldissemination.south -| lrevocation.south east) +(3pt,-3pt) coordinate (local-br);

\draw[box] (local-tl) rectangle (local-br);

\path (local-tl) +(-3pt,3pt) coordinate (neighborhood-tl);
\path (local-br |- ndissemination.south east) +(3pt,-3pt) coordinate (neighborhood-br);

\path[box] (neighborhood-tl) rectangle (neighborhood-br);

\path (neighborhood-tl) +(-3pt,3pt) coordinate (global-tl);
\path (neighborhood-br |- grevocation.south east) +(3pt,-3pt) coordinate (global-br);

\path[box] (global-tl) rectangle (global-br);

\node at (sensors -| lrevocation) {Local};
\path (lrevocation) +(0,-0.65) node {Cooperative};
\path (nrevocation) +(0,-0.65) node {Global};

\end{tikzpicture}
	\caption{Different scopes at which detection mechanisms can operate.}
	\label{fig:detection-scopes}
\end{figure*}
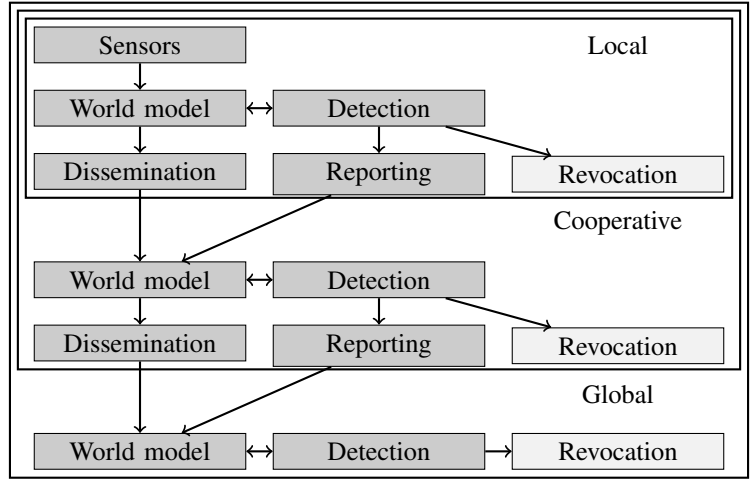

A first classification of detection mechanisms is by scope, as show in Figure~\ref{fig:detection-scopes}: local, cooperative and global detection.
Local detection refers to detection that checks internal consistency, and optionally the vehicles' sensors, as measures for correctness.
Cooperative detection relies on collaboration between vehicles (and possibly some \acp{RSU}).
Finally, global detection refers to detection that occurs at least partially with support from a back-end system.
This terminology is used throughout the literature describing the different mechanisms, although there is some disagreement regarding what an attackers capabilities in a cooperative setting are (refer to Section~\ref{sec:attacker-model} for more details).

Most detection mechanisms in the behavioral and plausibility classes operate purely locally, which makes them invariant to Sybil attacks.
However, this also makes it difficult to identify the attacker, or even the attack, due to the low amount of available information.
Some consistency-based mechanisms also operate locally, by comparing series of beacons from several vehicles, exposing them to Sybil attacks, in exchange for a much more fine-grained approach to detecting attacks.

On the other hand, there exist schemes that perform detection cooperatively: these are typically consistency- and trust-based detection mechanisms.
These detection mechanisms often rely on an honest majority and exchange messages between participants to detect inconsistencies or untrustworthy participants.
In particular the category of trust-based detection is based on cooperation between nodes.
Most schemes that operate as a reputation mechanism also incorporate a mechanism to perform reporting and/or revocation on the cooperative and global level.

\subsection{Taxonomy of Misbehavior Detection}
\label{sec:misbehavior-detection-taxonomy}
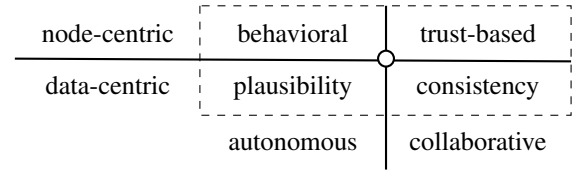
\begin{figure}

\centering
\begin{tikzpicture}
\matrix (m) [matrix of nodes,row sep=0, column sep=0,%
    nodes={minimum width=7em,minimum height=2em}]
{
node-centric & behavioral   & trust-based   \\
data-centric & plausibility & consistency   \\
             & autonomous   & collaborative \\
};

\coordinate (top) at (m-1-2.north east);
\coordinate (bottom) at (m-3-2.south east);
\coordinate (left) at (m-2-1.north west);
\coordinate (right) at (m-2-3.north east);

\coordinate (tl) at (m-1-2.north west);
\coordinate (br) at (m-2-3.south east);

\draw[thick] (top) -- (bottom) (left) -- (right);
\draw[thick,fill=white] (top |- left) circle (3pt);

\draw[dashed] (tl) rectangle (br);

\end{tikzpicture}

	\caption{Taxonomy of misbehavior detection.}
	\label{fig:misbehavior-detection-taxonomy}
\end{figure}

We have designed a taxonomy to categorize mechanisms detecting attacks within the attacker model discussed above, consisting of two classification aspects and four classes, as shown in Figure~\ref{fig:misbehavior-detection-taxonomy}.

The first aspect we use for classification distinguishes between node-centric and data-centric mechanisms, which has regularly been used in the literature.
For our purposes, node-centric mechanisms are defined as mechanisms that primarily concerned with the participants of the network.
For example, they can verify the forwarding behavior of a  node by analyzing packet frequencies, correctly formatted messages, and so on to decide on its trustworthiness.
Alternatively, they are focused on the interaction between participants, verifying their trustworthiness based on the correctness of previous messages.
Obviously the correctness needs to be tested some other mechanism: those mechanisms are usually data-centric in nature, meaning they use the content of the message to determine its validity, independent of who transmitted the message.
Data-centric and node-centric misbehavior detection are therefore mostly orthogonal, leading many authors to propose combinations of both types.

The second aspect we use to classify misbehavior detection mechanisms is the distinction between mechanisms that analyze messages from a single vehicle (autonomous) and mechanisms that attempt to deduce misbehavior from multiple vehicles (collaborative).
This distinction is best illustrated using an example from a data-centric mechanism.
Such mechanisms can analyze one or more messages from the same sender for invalid behavior (autonomous), or they can compare messages from different sources for consistency or deviations (collaborative)
A significant advantage of autonomous detection is that the mechanism will function independent of any attackers that may be present, while collaborative mechanisms rely on the fact that an honest majority exists.
Note that autonomous detection sometimes includes the use of the vehicles' own sensors (e.g., when determining the approximate source of a radio transmission, or the relative position).
Some consider autonomous approaches combined with vehicular sensors to be the only one that provides a reliable system.
However, the challenge with autonomous detection mechanisms is that they are often imprecise, and cannot necessarily detect intelligent attacks that use knowledge of the detection algorithm.
This imprecision is best exemplified by plausibility checks, such as a limit on the claimed speed that is accepted; an intelligent attack can always attack such checks by choosing values close to the limit.
To avoid a large number of false positives, the limit on the accepted speed should be high enough to deal with outliers (e.g., speeding vehicles), but low enough to detect malicious behavior.
If correctly employed, a set of plausibility checks bounds the message space available to the attacker, but also provides a space in which the attacker can choose messages that will be considered valid.
This resulting bound can be used by employing collaborative mechanisms to provide a higher quality of detection than is possible using either approach individually.
Indeed, the field is progressing towards these ideas, as evidenced by several detection mechanisms we discuss in this survey.
Having discussed these classifications, we now describe the resulting four classes from Figure~\ref{fig:misbehavior-detection-taxonomy}.

\subsubsection{Node-centric Misbehavior Detection}
\paragraph{Behavioral}
The first branch of node-centric detection is behavioral.
This type of detection exploits patterns in the behavior of specific nodes at a protocol level.
Information considered by these mechanisms includes the amount of messages transmitted by a node or the correctness of their format.
A core aspect of behavioral mechanisms is that their analysis is focused on a node-basis and typically does not consider data semantics, as done by data-centric mechanisms.
An example of a behavioral misbehavior detection mechanism is the concept of a Watchdog~\cite{Marti2000:Mitigating} that was introduced for the security of routing in mobile ad-hoc networks.
In a Watchdog, each node monitors the network to verify that its neighbors actually forward the messages they are supposed to forward.

\paragraph{Trust-based}
The other branch of node-centric detection is trust-based.
Trust-based mechanisms exploit the fact that many nodes in the network are honest, and that infrastructure is available to remove malicious nodes.
Trust-based detection includes reputation systems, which rate node behavior over time, but also voting schemes that allow vehicles to vote on the correctness of information.
Trust-based detection can occur either locally or with infrastructural support.
In the latter case, issues regarding privacy are a serious concern, which we address in Section~\ref{sec:pseudonyms}.
Trust-based detection often relies on input from other detection mechanisms to update the reputation of nodes in the network.

A core advantage of trust-based mechanisms is that they simplify the revocation process.
However, the ephemeral nature of \acp{cITS} poses additional challenges to trust-based mechanisms, particularly when determining their initial trustworthiness, as well as the information to update it.
Data-centric detection mechanisms or behavioral detection mechanisms may be used for this: the trust-based detection mechanism is then used to distinguish between legitimate and malicious behavior.
Both reputation- and voting-based mechanisms have to deal with Sybil attacks~\cite{Douceur2002-Sybil}, where an attacker abuses the reputation mechanism by creating multiple identities.
These attacks can be used to either amplify an attack or exclude legitimate nodes from the network, and is one of the core research challenges for trust-based misbehavior detection, especially in combination with privacy requirements.
Similarly, if an honest node suddenly starts misbehaving (e.g., due to a faulty sensor), this will not be detected quickly by reputation mechanisms.

\subsubsection{Data-centric Misbehavior Detection}

\paragraph{Consistency} 
Consistency-based detection uses relations between packets, typically from multiple participants, to determine the trustworthiness of newly received data.
For example, a consistency-based detection mechanism may consider a previously computed average speed of vehicles on a highway to judge newly received messages.
Messages that deviate from the average are inconsistent with the known state and can thus be considered suspicious.
Alternatively, pairwise comparison of messages from different vehicles is also considered consistency-based.
Consistency-based detection has the advantage that only limited domain knowledge is required to design reasonable schemes.
However, a local honest majority is often required to draw reliable conclusions: if several colluding attackers surround a victim, consistency might exclude information from legitimate vehicles.

\paragraph{Plausibility}
Plausibility-based detection uses a specific underlying model of data in order to verify if the transmitted information is consistent with this model.
For example, plausibility of movement can be verified from two subsequent beacon messages by examining the distance traveled between them and comparing it to the speed in these messages.
Plausibility-based typically allows for a very rudimentary but fast analysis of received packets.
It is performed by considering packets from individual senders.
The information in the packet is either tested against a prediction of the model, or the model is used to judge whether the information in the packet is a plausible next step according to the model.
Because there is an underlying model, the mechanism can directly express the plausibility of the message in a probability, which can be input for further misbehavior detection mechanisms.
The models used for plausibility vary from narrowly defined models like the laws of physics up to models that allow a wide range of variation, such as a model that predicts driver behavior.
The narrowly defined models can be effectively used to filter incoming packets for ``impossible'' messages that can be discarded directly.
A major advantage of plausibility-based detection is that the mechanisms are always applicable, even when an honest majority does not exist.
A significant disadvantage is that a model of the underlying data is required, and the utility of the mechanism depends directly on how accurate this model is.

\subsection{Impact of Pseudonyms} \label{sec:pseudonyms}
As mentioned previously, pseudonym systems can have an impact on the types of misbehavior detection that can be performed.
In particular, pseudonym schemes aim to hinder linkability of different information items.
Linkability is an important aspect of many models used for misbehavior detection, and sometimes even critical for safety applications.
We propose four simple classes that we will use to classify detection mechanisms in our survey: full, explicit, implicit and no linkability, which are briefly explained in the following.
For more details on the subtleties of pseudonyms, we refer interested readers to a survey by Petit~et~al.~\cite{Petit2014-Survey}.

\paragraph{Full linkability} means that all messages transmitted by the same OBU can be linked, i.e., there is no pseudonym change at all, and thus essentially no privacy.
All possible misbehavior detection mechanisms can be realized at this level and some types of misbehavior detection even require this level of linkability.
Examples of such schemes include trust-based approaches like OREN~\cite{Bilogrevic2011} and~\cite{Rawat2011-securing}.

\paragraph{Explicit linkability} refers to any type of linkability that allows direct access to an identity.
This means that linking is technically possible; however, there are several ways to implement this.
The issued certificate can either contain the \emph{identity in an encrypted form}, a \emph{direct mapping} from certificates to identity can be stored in the back-end, or this mapping may be split across multiple entities through \emph{organizational separation}.
Using each of these classes, it is technically possible to link pseudonyms, but this capability is limited by the pseudonym scheme.

\emph{Encrypting an identity} in the certificate is an attractive approach, because there are cryptographic schemes that enable conditionally revealing this identity (e.g., PriPAYD~\cite{Troncoso2011-PriPAYD}, or double spending prevention in electronic payment systems).
However, it is not clear whether this can also be applied for misbehavior detection, especially if revealing the identity should happen locally.
This system also requires that detection mechanisms are standardized, and in most cases part of the certification infrastructure, which is an unrealistic assumption.
Using a \emph{back-end mapping} is a solid alternative to this approach, in terms of implementation effort.
Here, each certification authority simply stores the relationship between pseudonyms and real identities, revealing malicious users as needed.
This is particularly useful when combined with direct back-end connectivity, either through \acp{RSU} or another technology, such as UMTS or LTE.
However, this requires full trust in the back-end system, as it can revoke the privacy of the users.
Another advantage is that a back-end could perform its' own misbehavior detection, and exploit the significantly higher computational power available to it.
An example of such a powerful scheme is~\cite{Bissmeyer2012}.
A middle way between these two types is \emph{organizational separation}, which also stores the link between two pseudonyms, but spreads the information necessary to reveal a user across organizational boundaries.
VToken~\cite{Schaub2010-VToken} and Butterfly Keys~\cite{Whyte2013-Butterfly} are different proposals that achieve this; overall it seems that industry and academia both prefer organizational separation over purely technical solutions.
The core advantage is that the organizational separation allows individual decisions on a case-by-case basis, if needed.
However, for misbehavior detection, this means that linkability between pseudonyms cannot be achieved locally in the vehicle, but only on an organizational level.
The advantage is that confirmed misbehavior can be acted upon immediately, and some mechanisms may be able to detect attacks by using a partial linking protocol to perform specific checks for messages that are suspicious if the pseudonyms would belong to the same identity (or vice-versa).

\paragraph{Implicit linkability} or \emph{inference} allows pseudonym linking even when no mechanism for direct identification or linking is available.
There are three different sources of information available for this purpose: \emph{certificates}, \emph{message content} and \emph{signal properties} during transmission.
Using this information, partial or complete identities may be derived, although there is no guarantee that such a linking approach will always be able to identify every vehicle.
For example, certificates typically contain attributes of a vehicle, such as length, height and color.
These properties allow full identification of unusual vehicles on the road, such as a yellow limousine, but not the identification of several black vans of identical make.
Although this is an example of linking based on certificate content, this type of limitation applies to all inference-based linking, and while it can be bounded by combining different types, it will have implications for misbehavior detection.

Secondly, implicit linkability does not fundamentally provide linkability through an identity, but through similarities in messages.
Implicit linkability only provides links between messages, which will allow a receiving vehicle to create 'pseudo-identities' for groups of pseudonyms that are considered to be the same vehicle.
This makes it more challenging to exchange information about particular vehicles with other participants, and can also make it difficult to verify misbehavior and revoke the misbehaving vehicle.
Nevertheless, many misbehavior detectors will need to function under these assumptions, as they provide the most realistic environment in which individual vehicles have to perform misbehavior detection (as opposed to detectors in the back-end).
Examples of inference-based systems include virtually all data-centric security mechanisms.
For signal-based inference,~\cite{Guette2007-sybil} and~\cite{Xiao2006-detection} are good examples, while for inference on message content, a good example is the Kalman filter~\cite{Stuebing2010-verifying,Stuebing2011-two-stage}.
Inference based on certificates is a relatively new idea that is sometimes captured within inference through messages, and is not commonly considered because it conflicts with the assumption that these certificates are intended to be pseudonymous or anonymous towards other participants.
Therefore, they are typically assumed not to reveal this information, making such detection mechanisms obsolete.

\paragraph{No linkability} is the idealized scenario in which it is impossible to determine whether two messages originate from the same or from distinct vehicles.
Although this results in maximum privacy, such a scenario makes it nearly impossible to provide any working \acp{cITS} applications, let alone securing them.
For example we can still attempt to perform misbehavior detection on individual messages.
The goal of such an analysis is to detect unusual or inconsistent messages, which either contain obviously incorrect values: this includes unrealistic values (speeds of 500m/s) and values that do not match the perceived scenario (speeds of 50m/s in a traffic jam).
The value of such detections is limited from a security point of view, but filtering malicious packets may still be beneficial for applications.
Beyond that, not much more is possible: at this level of linkability, the inability to detect Sybil attacks makes it extremely difficult, if not impossible, to build a structured detection mechanism.
Unfortunately, this means that there is a necessary trade-off between privacy as opposed to security and functionality -- necessarily, we are forced to sacrifice theoretically perfect privacy in order to gain functionality and security.
However, we remark that even in this scenario, tracking is possible by simply following a vehicle.
Because tracking vehicles by following them individually is unavoidable, it serves as a good baseline against which privacy can be tested.
However, detection schemes that require no linkability at all will definitely not negatively impact the privacy of the system.

\section{State of the Art}
\label{sec:state_of_the_art}

In this section, we present state of the art mechanisms for misbehavior detection in \acp{cITS}, as structured by our taxonomy.
Although some mechanisms in this survey are designed for related types of ad-hoc networks, such as \acp{MANET} and \acp{WSN}, we will focus on \acp{cITS}.
Separate surveys exist for these networks, though~\cite{Djenouri2005:MANETSurvey,Zhou2008-WSNSurvey}, and we will only use these as examples of how works for \acp{cITS} evolved from them.
After discussing the individual mechanisms for our state of the art, we provide a summarizing table of our analysis in Section~\ref{sec:tableoverview}.
This summary will include the additional parameters we discussed in Section~\ref{sec:misbehavior}; the scope of the detection and its compatibility with pseudonyms.

\subsection{Node-centric Mechanisms}

The idea of node-centric detection mechanisms is to use knowledge about the senders of messages in order to detect malicious senders.
Signatures are typically used to achieve this goal, with certificates provided by a PKI (cf. Section~\ref{sec:system_model}), that allow identification of a messages' sender.
Once the authenticity of a message sender is verified, it can be used to correlate messages originating from the same sender and thereupon analyze its behavior.
Such analysis can include whether the message frequency, headers and content are in line with protocol specifications.
Using past behavior, we can then assign a trust value to information sources based on previous interaction with them.
The assumption is that malicious messages most frequently originate from information sources that have misbehaved in the past.

In our taxonomy, we classify node-centric mechanisms as either behavioral or trust-based.
In earlier work for \acp{MANET}, node-centric mechanisms are the most popular, because these networks are focused on providing an abstract service to network participants, as opposed to a set of services that is inherently tied with data semantics, as in \acp{cITS}.
Node-centric approaches can be further subdivided by their specific approach; an overview of these approaches, with the example mechanisms we discuss in this survey, are shown in Table~\ref{tab:node-centric}.
We explain these types, as well as exemplary mechanisms, in the remainder of this section.

\begin{table*}
  \centering
  \begin{tabular}{llll}
    Type & Category & Examples\\
    \toprule
    Watchdogs & Behavioral &  Hortelano~et~al.~\cite{Hortelano2010}\\
    Flooding detection & Behavioral & Hamieh~et~al.~\cite{Hamieh2009}, Pu\~nal~et~al.~\cite{Punal2012-VANETs}\\
\midrule
    Central Sybil detection & Trust-based & Chen~et~al.~\cite{Chen2009-robust}, P$^2$DAP~\cite{Zhou2011-p2dap}, and Footprint~\cite{Chang2012}\\
    Node validation and reputation & Trust-based & LEAVE~\cite{Raya2007-Eviction}, OREN~\cite{Bilogrevic2011}, SLEP/PRP~\cite{Zhuo2009}, Sowattana~et~al.~\cite{sowattana_distributed_2017}\\
    Decision logics & Trust-based & Raya~et~al.~\cite{Raya2008}, Rawat~et~al.~\cite{Rawat2011-securing}\\
    Event validation and consensus & Trust-based & PoR~\cite{Cao2008}, z-smallest~\cite{Hsiao2011}, Leinm\"uller~et~al.~\cite{Leinmueller2010-its-congress}, CoE~\cite{Kim2010}, Petit~et~al.~\cite{Petit2011}\\
\bottomrule
  \end{tabular}
  \caption{Node-centric detection approaches} \label{tab:node-centric}
\end{table*}

\subsubsection{Behavioral mechanisms}
\label{sub:behavioral_mechanisms}

Behavioral mechanisms are focused on the behavior of a particular node.
This mainly concerns packet headers and meta-information like message frequency.
Behavioral schemes in \acp{cITS} typically focus on identifying nodes which send messages too frequently or nodes which modify the message content in a way that does not adhere to protocol standards.
As these attacks are not fundamentally different from attacks that some classes of network intrusion detection mechanisms aim at, there are not many \ac{cITS}-specific schemes available.
Behavioral mechanisms are especially popular to protect networks where routing attacks and fairness play an important role, such as \acp{MANET}.
Some of these misbehavior detection mechanisms have been adapted to work in \acp{cITS} scenarios.

Before discussing mechanisms specifically designed for \acp{cITS}, we take a brief historical perspective and discuss a seminal work developed for \acp{MANET}~\cite{Marti2000:Mitigating}.
This paper introduces two tools for misbehavior detection: the Watchdog and the Pathrater, which they evaluate for the multi-hop routing mechanism called \ac{DSR}.
The essence of the watchdog mechanism is that each node that participates in routing monitors the network after forwarding a packet to a next hop.
This node can then overhear whether the next hop forwards the packet or not, and therefore establish whether it is correctly behaving as defined by the protocol (in this case, \ac{DSR}).
Because a lossy channel might cause transmissions to be lost, a Watchdog should be configured with a threshold before it detects a node as malicious.
Challenges for this mechanism include loss or collision on the channel, as well as false reports generated by malicious or colluding nodes.
The aspect of colluding nodes is discussed in more detail in Section~\ref{sub:node_trust_based}, where we discuss trust-based mechanisms that deal with this issue.

\paragraph{Watchdogs} There are many proposals that adapt the original Watchdog mechanism for specific requirements of \acp{cITS}, because they have the advantage that they are fully independent of transported content.
However, they can clearly only be applied in settings where multi-hop communication occurs (either through routing or flooding-style approaches).
One such work is that of Hortelano~et~al.~\cite{Hortelano2010}, who set out to evaluate the usefulness of the traditional Watchdog mechanism for \acp{cITS}.
The core assumption is that routing is standardized, and vehicles can predict the (expected) behavior of their neighbors.
To accommodate packet collisions and noise on the wireless medium, the required number of re-broadcasts is lowered by a certain threshold (called \emph{tolerance threshold}), to reduce the number of false positives, similar to the \ac{MANET} case.
In addition, the importance of older results degrades more quickly over time, to account for the increased mobility (the authors call this \emph{devaluation}).
Once the watchdog detects malicious behavior, it is logged in a local file, but reports are not forwarded to other vehicles or a centralized instance.
Their evaluation shows that it is difficult to find a global threshold for deciding at which point misbehavior should be detected; setting this dynamically could lead to higher accuracy.
In addition, we note this paper does not address privacy in their analysis, making the true suitability for \acp{cITS} unclear at best.
As pointed out by the authors, several vulnerabilities remain; additionally, there is no protection against Sybil attacks.
We note that in our more general attacker model, the attacker could use selective jamming in addition to further amplify several attacks.

\paragraph{Flooding detection} In addition to mechanisms that are designed to detect attacks on routing, a very useful mechanism involves the detection of a related type of attack: flooding.
Although Watchdogs are capable of detecting some types of flooding, we create a separate category for detection algorithms that detect other types of flooding, such as jamming and MAC attacks.
In contrast to Watchdogs, Hamieh~et~al.~\cite{Hamieh2009} describe a detection mechanism for jamming attacks based on detecting patterns in radio interference.
The assumption is that a jamming attacker behaves intelligently, in order to gain an advantage through his jamming.
This approach, known as selective jamming, is a common technique to avoid being easily detected due to constant jamming of the wireless channel.
The proposed approach makes use of the fact that a selective jamming attacker will wait until regular transmissions occur until they jam the wireless medium.
Hence, a correlation coefficient between correct reception time and time where errors occur is calculated.
If the correlation is high, that is, if the medium is jammed most of the time when regular reception should occur, the medium is considered jammed.
In order to achieve useful results, the authors took into account realistic reception and error probabilities as a baseline.
Only if the correlation is unusually high, the medium can be considered jammed.
The proposed method is interesting because the correlation can be passively calculated with a simple formula, and because detection of selective jamming is an important problem that is often neglected in security-related works.
A deeper discussion and classification of jamming attacks is provided in a later work by Pu\~{n}al~et~al.~\cite{Punal2012-VANETs}, which discusses different types of jamming attacks and analyzes their feasibility.
In particular, they show some types of jammers are capable enough to prevent communication altogether with high probability, illustrating the necessity of jamming detection.

\subsubsection{Trust-based mechanisms}
\label{sub:node_trust_based}

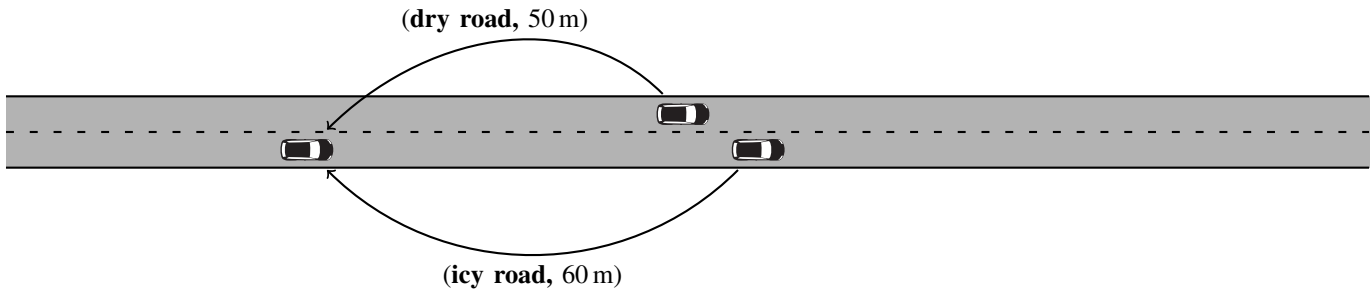
\begin{figure*}[ht]
\begin{tikzpicture}
\road
\path (lane1-left) ++(3,0) node (d) {\car};
\path (lane2-left -| d) ++(5,0) node (s1) {\car};
\path (d) ++(6,0) node (s2) {\car};
\draw[->,thick] (s1) edge[out=135,in=45] node[above] {(\textbf{dry road,} 50\,m)} (d);
\draw[->,thick] (s2) edge[out=-135,in=-45] node[below] {(\textbf{icy road,} 60\,m)} (d);
\end{tikzpicture}
\caption{Example showing conflicting event notifications.}
\label{fig:conflicting-info}
\end{figure*}

Similar to behavioral mechanisms, many trust-based mechanisms for \acp{cITS} are rooted in mechanisms that were developed for \acp{MANET}.
These partially evolved from mechanisms such as the Watchdog~\cite{Marti2000:Mitigating}, which provide metrics to establish the trustworthiness of a node.
To aggregate this trust, distribute it among nodes, and provide it to a back-end system, a mechanism is required that not only filters malicious nodes as quickly and efficiently as possible, but also prevent attacks on the mechanism itself.
For example, Pathrater~\cite{Marti2000:Mitigating} aggregates the Watchdog results, but it may be attacked through Sybil attacks (as the authors also discuss).
Core issues for trust-based mechanisms are Sybil attacks on the one hand, and high mobility and brief connectivity on the other.
These challenges are much stronger in \acp{cITS}, as the connectivity between vehicles is sporadic, and privacy requirements lead to a vehicle being allowed to use multiple identities.
This is especially significant when conflicting information is received from two sources that are either equally trustworthy, or both unknown up to this point, as illustrated in Figure~\ref{fig:conflicting-info}.
To account for these challenges, authors often propose the use of data-centric mechanism output, which they then integrate into the trust-based mechanism; we will discuss these approaches towards the end of this section.
We include these mechanisms here, because the primary focus is the development of the trust management, and not the data-centric detection process.
There are also dedicated surveys for trust schemes to be found in the literature, such as the work by Rivas~et~al.~\cite{Rivas2011}.

\paragraph{\ac{RSU}-supported Sybil attack detection} There are many schemes that operate using a centralized authority to detect Sybil nodes.
The approaches we describe here are based primarily on interactions between infrastructure and vehicle.
By grouping vehicles based on the \acp{RSU} they interact with, the assumption is that the Sybils of an attacker will always follow the same interaction path as the attacker.
We describe several of these schemes here, which are developed independently, but have similar properties: trajectory signatures~\cite{Chen2009-robust}, P$^2$DAP~\cite{Zhou2011-p2dap}, and Footprint~\cite{Chang2012}.

\cite{Chen2009-robust} points out that Sybil nodes that originate from the same vehicle will always have unrealistically similar trajectories over time.
With this observation, they design a protocol to request special signatures, obtained from \acp{RSU} by each vehicle, which can be used to build a trajectory.
Whenever determining trustworthiness, the identities with identical recent trajectories are assumed to be the same vehicle, and are combined into a single identity for this step, evening out the effect of Sybil attacks.
Although this approach is theoretically appealing, it relies on a variety of assumptions most significantly; a widespread deployment of \acp{RSU}, which cannot be assumed everywhere, and full trust in all \acp{RSU}.
There is also the practical issue of bandwidth overhead versus distinguishability; the larger the amount of signatures $r$, the more expensive the protocol.
This also potentially enables denial of service attacks, because each request requires a much larger response (i.e., the signatures).
Next, some of the values may be set to the value \textit{None} for time slots in which no signature was received from an \ac{RSU}, which could be used to cheat the protocol.
Finally, any vehicle will completely lose privacy if they want to prove that they are not Sybil nodes, because they must reveal their position trace (i.e., the set of $r$ signatures).

In P$^2$DAP, Zhou~et~al.~\cite{Zhou2011-p2dap} proposes a baseline to verify all messages, which constitutes a relatively specific attack.
Whenever a triplet of time, location and event type is signed by the same vehicle with different pseudonyms, it is considered an attack.
In order to prevent these attacks, the authors propose an inherent linking between pseudonyms based on hash functions.
A semi-trusted \acp{RSU} is responsible for checking whether an attack occurs using this linking, and reports it to the central authority, which can resolve pseudonymity completely.
By using two classes of pseudonyms, the \acp{RSU} can only determine that two pseudonyms belong to the same class (called coarse-grained group), which corresponds to a vehicle, and it cannot directly identify that vehicle.
Although this scheme provides relatively strong security in this setting, being able to link arbitrary pseudonyms is a huge privacy issue.
Apart from that, the central authority has complete knowledge, which is problematic.
Another problem with this scheme is that the event types, time and spatial granularity must be standardized and require a single central authority, which is not practical in a real world where border crossings are common.

Footprint~\cite{Chang2012} improves on Chen~et~al.~\cite{Chen2009-robust}: it also exploits central authorities by using similarity of trajectories, which are generated using signed messages by \acp{RSU} that a vehicle passes.
In Footprint, these trajectories are cryptographically protected, and consist of special signatures, requested by the vehicle from the \acp{RSU} is has seen while driving.
Footprint works by bounding the potential set of valid distinct trajectories an attacker can create.
This bound is, in the worst case, the power set of trajectories, but can be limited in size using a test (which we do not discuss in detail here).
The authors assume that real trajectories are sufficiently distinct; by forcing the attacker to obtain signatures through the \acp{RSU}, the bound is created based on the real path of the attacker.
Then, when detecting Sybil attacks, all trajectories that are suspiciously similar are considered as coming from the same vehicle (referred to as a Sybil community).
The authors use the trajectories for every message as an authentication mechanism, which allows any vehicle to compute the Sybil communities and avoid Sybil attacks.
The signatures of \acp{RSU} are time-dependent and unpredictable, which means that location privacy is achieved against long term tracking.
The other disadvantages remain, however; wide \ac{RSU} deployment is still needed.

\paragraph{Node validation and Reputation} Following are multiple papers specifically aiming at mechanisms that use explicit voting.
Here, voting is used to decide which nodes are (not) trustworthy.
These are distinct from consensus schemes, which use identities to vote on whether or not a claimed event is true; such schemes are discussed below.
Voting mechanisms strictly require protection against Sybil attacks, and those attacks are thus usually not part of the attacker model.

Raya~et~al.~\cite{Raya2007-Eviction} have also been among the first to present a system for locally evicting nodes, including the possibility to perform global revocation as a result using a protocol called LEAVE.
Here, vehicles exchange accusations about potential attackers; as soon as a threshold is reached, a vehicle is evicted temporarily.
The set of accusations can be distributed in an aggregated message, which can be used to immediately ignore a vehicle, and used as a tool for global revocation once it reaches the CA.
A core advantage of this approach compared to reputation systems is reduced detection latency, because trust does not need to be built over time in LEAVE.
The authors argued that local eviction is especially suitable for vehicular networks because of the low communication overhead and quick reaction time to attacks compared to global revocation.
However, global revocation based on analysis of the collected local reports is foreseen as an orthogonal countermeasure against persistent attackers.
Some disadvantages of the scheme is that it may be vulnerable to Sybil attacks and may have privacy issues, depending on the type and implementation of the pseudonyms that are used.

Similarly, Zhuo~et~al.~\cite{Zhuo2009} aims to remove misbehaving insiders from the local neighborhood or the network.
To achieve this goal, two specific mechanisms are proposed: suicide-based local eviction (SLEP) and permanent revocation (PRP).
As their names suggest, SLEP evicts attackers locally and PRP achieves global revocation of misbehaving vehicles.
Both protocols can be applied even if pseudonym schemes are used in order to preserve user privacy.
The local eviction protocol works using a so-called suicide mechanism, which is used to discourage false accusations.
The assumption is that vehicles in the direct neighborhood of an attacker are able to detect bogus messages, for instance, by comparing them to local sensor information about the same event.
If a vehicle detects an attack, it broadcasts a message accusing the potential attacker vehicle.
Surrounding vehicles receiving the message will henceforth ignore messages from the accused vehicle, but they will also ignore further messages from the \emph{accusing} vehicle.
To prevent multiple honest vehicles from committing suicide in response to the same misbehavior, a random back-off timer is employed before disseminating accusations.
To achieve global revocation, each vehicle periodically reports its local blacklist of possible attackers to the back-end, which then uses the trust level of each accuser to decide on revocation.
Unlike Raya~et~al., the authors specifically address the use of pseudonyms, and SLEP relies on it to allow vehicles to ``re-join'' the network after a successful accusation.

Raya~et~al.~\cite{Raya2010} and Bilogrevic~et~al.~\cite{Bilogrevic2010,Bilogrevic2011} have presented closely related game-theoretic approaches to combine their eviction scheme from~\cite{Raya2007-Eviction} with a suicide mechanism similar to that of Zhuo~et~al., combining the previous approaches.
Here, a suicide action results in immediate local revocation of both accuser and accused, but vehicles have the alternative to vote by transmitting an accusation, or abstain from the protocol completely.
Each course of action is associated with a certain cost, where suicide has the highest cost.
In addition, cost values take into account the fact that abstaining from any action for longer periods of time will leave the attacker more time to do harm to the network.
Using game theory, each vehicle decides when and whether to take which action in order to minimize the cost for all participants using a variety of strategies.

Criticizing a number of existing revocation schemes including the aforementioned papers, Liu~et~al.'s~\cite{Liu2010} goal is to identify the limits of such schemes in vehicular networks.
One of the main points argued in the paper is that a local honest majority, an assumption made by the revocation schemes discussed so far, is not as likely as the previous authors argued.
If an attacker manages to create a local majority, she can create false accusations and falsely remove honest vehicles from the local communication network.
If pseudonyms are used to protect user privacy, the authors considered it feasible for an attacker to use multiple pseudonyms in parallel to create such a local majority.
A general conclusion for all voting-based schemes is thus that parallel use of multiple pseudonyms should be prevented by the underlying pseudonym mechanism used.
Moreover, Liu criticized Raya~et~al.'s game-theoretic approach discussed above, arguing that the assumption that each vehicle acts in it's own interest might be flawed.
The reasoning is that the software running on the vehicles is not programmed by the vehicle owner herself, but by the manufacturer.
The authors note that manufacturers may optimize their own cost, which could invalidate the game-theoretic results that were obtained by assuming that all vehicles optimize their individual costs.

Sowattana~et~al.~\cite{sowattana_distributed_2017} discuss a voting-based Sybil detection scheme.
This scheme has similarities with earlier witness/verifier schemes proposed by~\cite{Xiao2006-detection} among others.
Similar to those schemes, each receiver validates the validity of received beacons by transmission range of those neighbors.
However, in contrast to earlier work, this scheme generates trust on specific vehicles through a voting scheme based on these results.
The voting process is the focus of this work, and the authors evaluate the performance of this scheme using state-of-the-art methods.
However, the voting process is itself vulnerable to attacks, and those attacks are not analyzed in this work.
It remains to be seen whether the limited space of positions for Sybils is sufficient to negate attacks on this voting scheme.
Privacy is not covered in this work at all, but we assume pseudonyms are used to generate Sybils.

\paragraph{Decision logic} To establish the trustworthiness of vehicles, which is used as input by the reputation mechanisms discussed above, authors often use a decision logic.
This is similar to traditional trust network approaches, which are typically adopted for this purpose.
Raya~et~al.~\cite{Raya2008} present one such adaptation, which uses both data-centric and node-centric information to establish trust in specific nodes.
However, the final decision is made by deciding on the overall trustworthiness of a node, rather than deciding the individual trustworthiness of messages.
The authors used a combination of three different factors to evaluate trust in nodes.
This includes default trustworthiness, which is based on the type of certificate (e.g., police cars); event- or task-specific trustworthiness, which matches the type of vehicle to the event; and dynamic trustworthiness, which captures message-specific information like proximity to the event.
Once each factor is computed, it is combined with the data-centric information and only then does the decision logic decide whether or not to trust the given node.
The authors evaluated a number of different decision logic implementations, but stated that no single mechanism performs best in all simulated configurations.
However, the Dempster-Shafer inference~\cite{Shafer1976,Dempster1976} was identified as the most promising technique.

Similarly, Rawat~et~al.~\cite{Rawat2011-securing} present an approach using Bayesian logic to predict the trustworthiness of vehicles.
Basically, they compute the maliciousness probability for a vehicle $i$ for a time $t$ given some observation $O_t$.
Their approach relies on Bayesian reasoning, i.e., computing $P(T_i \text{malicious} | O_t)$ by applying Bayes' theorem.
Unfortunately the authors do not specify how they obtain the necessary conditional probabilities (such as the probability of a message given previous observations).
The scheme requires pre-configured knowledge on the reception probability of a particular message.
Another issue is that the authors' model does not \emph{forget}: they always compute probabilities over the entire observation history.
They do point out their detection threshold will change over time, but do not specify how.
In their evaluation, they contrast signal-to-noise ratio with their computed trust values, based on a model of signal-to-noise ratio over relative distance and receive power.
This makes computation of the associated probabilities easier: if the claimed relative distance is falsified, their scheme can reliably detect this based on their model.
Their results are difficult to generalize, and even for this specific use case rely on a fixed transmission range for all nodes (including the attackers).

\paragraph{Event validation and consensus} A common challenge in trust-based mechanisms is the application of trust over multiple hops.
Event validation addresses this question by allowing each vehicle to use their identity as a vote in support of or against a specific event.
Similarly, consensus mechanisms do the same, but often relate more closely to data-centric mechanisms (e.g., agreeing on specific data, rather than events, or including data-centric detection results).

The work by Cao~et~al.~\cite{Cao2008} determine the correctness of event reports through voting. 
The core contribution of the paper is an efficient way to collect signatures from a sufficient number of witnesses without causing too much bandwidth overhead on the wireless channel.
In the proposed scheme, growth codes are used to achieve an efficient dissemination of the witnesses' signatures on the claimed events.
Growth codes are a type of erasure code, and are used to efficiently XOR together signatures over an unreliable channel for efficient collection.
They are optimized to decode as many signatures as quickly (in terms of transmission rounds) as possible; each of these signatures certifies a claimed event, and only events with many valid signatures are accepted as valid.
Growth codes give a relatively strong confidence, but introduce a significant delay before convincing a receiver of the validity of the message, because a threshold amount of signatures must be decoded first.
A strong criticism of this approach is that false negatives may cause significant problems; if insufficient signatures are received, events may be missed completely.
In addition, the attackers are assumed to have exactly one key pair each, which means the scheme is not compatible with pseudonyms.

Hsiao~et~al.~\cite{Hsiao2011} present a similar approach to determine whether a claimed event has actually occurred.
In order to prevent attacks, the senders collect a number of witnesses for each possible event.
For space efficiency purposes, $z$-smallest probabilistic counting is used, reducing the required amount of signatures that need to be attached to the message.
The idea of $z$-smallest is that, given $n$ elements uniformly distributed between $0$ and $1$, the $z$-smallest element gives an approximation of $n$ by calculating $\frac{z}{c}$, where $c$ is the value of the $z$-smallest element.
To protect against inflation, where attackers increase the number of witnesses of an event, each vehicle signs a hash of its vehicle id, the event type, location segment, and time of the event.
Only the $z$-smallest signatures are kept with the aggregate, meaning it is hard for the attacker to produce these $z$ signatures (because the hash is cryptographically secure).
There is no deflation protection in this scheme; the attacker can reduce the amount of signatures attached to the message.
The authors argue that an attacker will only try to produce fake events; there are easier ways to hide events that this mechanism cannot protect against anyway.
Similar to the previous paper, it is a hard requirement that vehicles are restricted to exactly one pseudonym per time period as the attacker model does not include Sybil attacks.

Leinm\"uller~et~al.~designed, amongst other mechanisms, a cooperative trust mechanism~\cite{Leinmueller2010-its-congress} to improve the results of their mechanisms by exploiting trust developed between vehicles.
The goal of the mechanism is to defend against attacks performed by static roadside attackers.
This is done by combining the data-centric minimum distance moved mechanisms with transitive trust.
Each vehicle evaluates locally whether sensor information implies that the sender of a message has moved a certain minimum distance.
Then, each vehicle broadcasts the collected information, indicating which neighbors passed the local the minimum distance moved check.
The assumption is that if a vehicle $A$ receives positive feedback about a vehicle $C$ from vehicle $B$ and $A$ has earlier identified $B$ as trustworthy, then $A$ will also trust $C$ without directly verifying it.
The scheme also incorporates neighbor information, which is shared among vehicles using periodic messages (proactive), although the authors also note that reactive mechanisms may be possible.
An honest majority assumption allows this mechanism to work, but only when the attacker is (mostly) stationary, which is a fairly weak attacker model.

The authors in~\cite{Petit2011} have proposed a scheme to build consensus about certain events in vehicular networks using a dynamic threshold for their consensus mechanism.
The proposed scheme collects a number of reports about the same event from surrounding vehicles until a certain threshold of supporting reports is passed, after which the message is considered to be trustworthy.
To cater for safety-relevant applications, the authors took into account a maximum waiting time, before which a decision must be reached.
For instance, in case of an accident warning message, the decision whether to trust the warning must be made early enough, so that the vehicle can still brake in time to avoid further collisions.
A main contribution of the authors is to implement the acceptance threshold in a dynamic way.
The threshold is adapted based on factors like distance to the claimed event and number of vehicles currently in the neighborhood.
However, like most consensus mechanisms, this approach suffers from potential Sybil attacks.
The dynamic threshold allows the mechanism to deal with density variations on the road, but as the authors also remark, selecting an appropriate initial value is still an open challenge.

Kim~et~al.~\cite{Kim2010} combine several data-centric and node-centric parameters of messages into a single curve, called the \ac{CoE} curve.
The validity of messages is determined by checking messages against the vehicles' own sensors, messages from other vehicles, and, where available, validation by infrastructure, making it a consensus mechanism.
This is combined with a relevance metric (i.e., whether the vehicle will ever interact with the event): only relevant, valid events are presented to the driver.
Validity is defined using a threshold curve, defining the required supporting evidence (similar in functionality to Petit~et~al.'s dynamic threshold).
This allows for fast responses in emergencies with less certainty, while it reduces the false positives for events where more evidence can be gathered over time.
We note that the \ac{CoE} could be vulnerable to Sybil attacks, depending on the implementation of the various sources of information that are specified.
In addition, this scheme is very difficult to generalize to other types of applications, because specific locations are required for events.
The example application the authors use, the \ac{EEBL} application, provides such locations, but it is unclear whether the scheme can deal with multiple lanes and urban settings, where there may be some uncertainty about the driver's path.
However, the \ac{CoE} is a powerful concept that can be employed to allow for maximum certainty before warning the driver, reducing false positive and negative rates.

\subsection{Data-centric Mechanisms}\label{sec:data-centric}

In this section, we consider the different forms of data-centric misbehavior detection that were identified in Section~\ref{sec:misbehavior-detection-taxonomy}.
Recall that the core distinction between consistency and plausibility is that consistency analyzes messages from different senders, while plausibility verifies messages from the same sender.
Plausibility mechanisms include analyzing physical layer signals to detect that they indeed come from the same source.
As before, we provide a brief overview of this section in Table~\ref{tab:data-centric}.

\begin{table*}
  \centering
  \begin{tabular}{llll}
    Type & Category & Examples\\
    \toprule
    Signal-based plausibility & Plausibility & Guette~\&~Ducourthial~\cite{Guette2007-sybil}, Ruj~et~al.~\cite{Ruj2011}, Xiao~et~al.~\cite{Xiao2006-detection},\\
     & &\hspace{1em} Sun~et~al.~\cite{Sun2017}, Yao~et~al.~\cite{yao_multi-channel_2018}\\
    Multi-rule plausibility & Plausibility & Leinmüller~et~al.~\cite{Leinmueller2008-decentralized}, PVN~\cite{Lo2007}, VEBAS~\cite{Schmidt2008}, Yan~et~al.~\cite{Yan2008-providing}\\
    Position prediction & Plausibility & Jaeger~et~al.~\cite{Jaeger2011-novel,Stuebing2011-two-stage}, Bißmeyer~et~al.~\cite{Bissmeyer2012-Particle}, Barnwal~et~al.~\cite{Barnwal2012-Zone}\\
    Vehicle dynamics & Plausibility & Yavvari~et~al.~\cite{yavvari_vehicular_2017}\\
    Post-event validation & Plausibility & Ghosh~et~al.~\cite{Ghosh2010-detecting}\\
\midrule
    Raw data comparison & Consistency & Bißmeyer~et~al.~\cite{Bissmeyer2010}, Zaidi~et~al.~\cite{Zaidi2016}, Grover~et~al.~\cite{Grover2011-sybil-attacks}, Leinmüller~et~al.~\cite{Leinmueller2008-decentralized}\\
    World modeling & Consistency & Golle~et~al.~\cite{Golle2004}, Dietzel~et~al.~\cite{Dietzel2010}, Bißmeyer~et~al.~\cite{Bissmeyer2012}\\
    Rule mining & Consistency & VARM~\cite{Rezgui2011-detecting}\\
    Machine learning on message contents & Consistency & Grover~et~al.~\cite{Grover2011-machine-learning}, Ghaleb~et~al.~\cite{Ghaleb2017}\\
    Sensor sharing detection & Consistency & Yan~et~al.~\cite{Yan2011-general}\\
\bottomrule
  \end{tabular}
  \caption{Data-centric detection approaches (some papers propose more than one type).} \label{tab:data-centric}
\end{table*}

\subsubsection{Plausibility}\label{sec:plausibility}
\label{sub:plausibility}

Plausibility checks can be used to quickly and efficiently filter packets that are malicious.
Typically simple instances of these mechanisms are assumed to exist by node-centric schemes to provide a way to determine trustworthiness of nodes.
However, plausibility checks can also be used as a more advanced tool to determine a numeric plausibility value, rather than just filtering out bad packets.
For example, one can analyze the speed or location of a vehicle over time, a receiver can identify its path and attempt to identify suspicious paths.
Plausibility checks are often used to detect attacks that involve Sybil nodes.
This relies on the fact that an attacker generating these Sybil nodes transmits the associated messages from approximately the same location.

\paragraph{Signal-based plausibility} One of the first applications for data-centric misbehavior detection was the identification of falsified positions by exploiting channel properties.
By detecting the source direction of the signal and using time-of-flight, vehicles can attempt to verify the position contained within a \ac{CAM}.
An overall challenge of these approaches is that the error of measurements outside a controlled environment are very high.

The authors of~\cite{Guette2007-sybil} provide a detailed analysis of Sybil attack detection through analysis of physical layer properties.
They assume that antennas, gains and transmission powers are fixed and known to all users of the \ac{cITS} (except for the attackers').
By applying signal models, they use the received signal strength to determine the approximate distance to the sender and apply this to verify the GPS position transmitted in each beacon message.
They show the theoretically possible areas where an attacker can transmit to cause the receiver to observe the desired received signal strength, in order to correspond to the received signal strength.
The authors also analyze the effect of using different antenna models (bi-directional and omni-directional) for the receiver.
As the authors point out, they do not consider special propagation models or GPS errors.

Similarly,~\cite{Ruj2011} exploits the relation between location, time and transmission duration for both beacons and event messages.
Time of flight is used to verify the positions transmitted by each sender, computing the time spent in the air by the message.
They assume light speed, i.e., $t_2=t_1*\frac{dist(l_{i, t_2}, l_{j, t_1})}{c}$, with $t_1$ as the claimed time of transmission, $t_{j, t_1}$ as the claimed location of sender $j$ and $l_{i, t_2}$ as the location of receiver at the time of reception $t_2$.
Although this approach is immune to Sybil attacks, it does not compensate for GPS errors, MAC delays or limited time synchronization between vehicles.
The authors also propose a post-event validation approach that looks at messages from other vehicles to verify specific event messages, which relies on correct positions using this time of flight approach.
Finally, we note that the authors consider a pseudonym change after a claimed event is malicious, which is detected through a lack of subsequent beacon messages from the same source.
This scheme is easily exploited to revoke legitimate vehicles by an attacker with even limited jamming capabilities.

Sun~et~al.~\cite{Sun2017}~propose a vehicle tracking scheme based on extended Kalman filters (EKF), a generalization of Kalman filters for non-linear systems.
The authors assume that EKFs are used to combine angle of arrival (AoA) and Doppler Speed measurements of the received signal to estimate the behavior of the sender.
By combining this information with the position information in the message, the estimation error of the EKF will not diverge, unless the vehicle misbehaves by transmitting false positions.
This is detected by applying a chi-square test to determine whether the $n$ most recent measurements reflect Gaussian behavior.
When the chi-square test detects a deviation, AoA and Doppler shift measurements from neighbors are used to estimate the position, which is then used to track the misbehaving vehicle despite the misbehavior.
The proposed scheme is discussed in highway scenarios only, but can deal with any attacks, regardless of collaboration.
Privacy is not addressed in this work; however, we suspect that the author's system can be deployed to track pseudonyms, while applying the work by St{\"u}bing~et~al.~\cite{Stuebing2010-verifying} to track vehicles across pseudonyms.

Sch{\"a}fer~et~al.~\cite{Schaefer-WiSec2016}~discuss how the Doppler effect can be used to validate movement in an approach similar to those discussed above.
Although their work primarily focuses on flying vehicles (e.g., airplanes), where distances are much greater and thus signal propagation takes much more time.
Thus, they also briefly address the application of their results to vehicular networks.
Specifically, they describe that because the required precision cannot be achieved in these networks at reasonable cost, they conducted experiments that measure Doppler Shift in sound.
Despite encouraging results with an equal error rate of zero, the authors state that approaches based on their scheme have two fundamental limitations.
These are attacks from moving vehicles and attacks where multiple antennas are used by the attacker.
Moving vehicles are unlikely due to the extreme velocities required, but multi-antenna attackers remain an open issue, which is the case for most schemes in this section.

Yao~et~al.~\cite{yao_multi-channel_2018}~provide an extensive analysis of RSSI-based schemes to detect Sybil attacks.
First, the authors describe an experimental setup that is used to make observations on how RSSI measurements change over time, and how these can be applied to detect attacks.
Basically, the authors state that one cannot directly apply RSSI measurements, and using a propagation model to estimate validity of messages through the RSSI is unlikely to provide reasonable results.
However, they also identify similarities between the RSSI of an attacker and its Sybil nodes over time, and discuss the use of similarity measures for these time series.
In their work, they use the dynamic time wrapping as a measure, which is particularly suitable because it can deal with message loss.
They also include Z-score normalization of all RSSI time series to prevent the attacker from simply using different power levels for each Sybil node.
Their work shows that even though RSSI itself is considered unreliable, it can be used to identify attacks when used as a time series.
One weakness of this system is that it breaks as soon as an attacker uses more than one radio.

\paragraph{Multi-rule plausibility} Another early idea was to use several sources of information to verify message validity, rather than just using signals to verify a position, which was sketched in~\cite{LHSW04:IDS_VANET}.
The authors introduce the concept to use on-board vehicle sensor data and data from different layers of the communication system to build a world model, with which newly received information can be compared.
Many of these multi-rule mechanisms also include some form of other detection mechanism, but often the focus lies on taking advantage of the many fast and simple checks that can be performed through plausibility checks.

Lo~\&~Tsai~\cite{Lo2007} discuss a type of data injection attack called the illusion attack, where the attacker injects false information into the \acp{cITS}.
To protect against this attack, the authors propose a plausibility validation network (PVN), which consists mainly of a checking module and a rule database.
The rule database contains a set of rules that govern whether certain information should be considered valid or not, by analyzing the individual message fields.
The authors provide a list of such rules to detect fake vehicles, which includes dropping of duplicate messages, the position being in transmission range, valid time stamps and others.
The set of rules is different per message type, and is default-deny (i.e., all rules must accept the message).
A weakness of this work is that it assumes that the attacker can only manipulate messages indirectly (i.e., through sensor manipulation, the attacker has no key material).
Because the rule database is shared amongst all participants, the attacker can generate only valid messages to avoid detection.

In~\cite{Leinmueller2008-decentralized}, the authors discuss a variety of data-centric mechanisms, including what they refer to as autonomous verification, which is similar to the PVN from Lo~\&~Tsai.
These checks were previously studied in~\cite{Leinmueller2006-improved}; we discuss the consistency mechanisms in the next section.
The checks introduced here are called acceptance range threshold (ART), mobility grade threshold (MGT) and maximum density threshold (MDT).
ART provides a very simple check that discards messages with position claims that are far outside the communication range of the receiver, and thus likely contain a falsified position.
MGT checks the distance moved between two beacons for suspiciously high speeds.
MDT examines the maximum amount of vehicles in a particular area; when too many beacons are sent from one area, the vehicle ignores further beacons from the same area, in order to limit the impact of Sybil attacks.
In addition to these checks, the authors mention the possibility of map-based verification and position claim overhearing.
Map-based verification assigns a plausibility value to the received beacons by comparing the location to a road map.
Position claim overhearing can be applied in (geo)routing scenarios: by comparing different overheard packets and their destinations, overheard packets can provide indications of a false position in the past.
All of these checks do not perform particularly well individually, as also discussed in~\cite{vanderHeijden2016}, but the potential for these checks lies in their efficiency, similar to the PVN.

Vehicle behavior analysis and Evaluation scheme (VEBAS), was presented by Schmidt~et~al.~\cite{Schmidt2008} as an attempt to build a full system up out of behavioral, plausibility and trust-based mechanisms.
VEBAS allows the local detection of unusual vehicle behavior.
Each vehicle analyzes all messages received from neighboring vehicles to detect possible misbehavior.
Schmidt~et~al. analyze the content of messages using different modules.
In this, most mechanisms are plausibility checks, with either positive (meaning the mechanism states the information is correct) or negative (information is false) results.
Some mechanisms analyze the behavior of a vehicle in the classical sense (i.e., behavioral mechanisms) while other mechanisms also make use of physical models.
As an example of behavior-based mechanisms, the authors used a maximum beaconing frequency threshold.
If a vehicle sends messages with an unusually high frequency, the behavior can be regarded as possible attack.
The other mechanisms discussed by the authors are plausibility-based mechanisms, some of which are re-used from earlier work.
Finally, the authors show a way to combine the output of the different detection mechanisms, the authors proposed to use the exponentially weighted moving average (EWMA) method.
Here, older information is integrated with less weight than fresh information, and different weight can be given to different modules.
This includes the possibility to send and receive reports from other nodes in the network (which adds a trust-based element to the mechanism): once a vehicle has accumulated a defined amount of information about surrounding node behavior, it will broadcast the results within the 1-hop neighborhood.
However, these recommendation reports are not trusted blindly by receiving nodes to prevent attackers from broadcasting fake positive behavior reports.

Multi-rule plausibility has also been used to detect attacks within cell- or cluster-based vehicular networks.
Although clustering itself has numerous disadvantages, verification approaches that analyze a leaders' behavior often require data-centric operation in order to exclude attacks.
Yan~et~al.~\cite{Yan2008-providing} present one such scheme where attacks are largely prevented by centralizing information at the leader.
The leader is responsible for announcing its cell member's positions, and the authors design a scheme where leader and cell members verify each other.
The local verification primarily relies on a comparison of GPS and radar measurements.
For additional accuracy, the authors introduce Sybil attack detection by using cosine similarity and two alternative information sources that are essentially consistency schemes.
Finally, the authors include a position prediction approach as one of their plausibility rules.
Nevertheless, the reliance on clustering makes this scheme particularly poor for privacy, and potentially reduces performance.

\paragraph{Position Prediction} As discussed in Section~\ref{sub:node_trust_based}, the advantages of node-based mechanisms are strengthened when pseudonyms can be resolved, or at least linked, in the local vicinity of a vehicle.
Position prediction essentially uses such techniques to predict behavior of vehicles and verify whether they follow an expected pattern.
Although the Kalman filter is the most common approach, alternatives include the use of particle filters, as discussed by Bißmeyer~et~al.~\cite{Bissmeyer2012-Particle}, and zone prediction, as discussed by Barnwal~et~al.~\cite{Barnwal2012-Zone}.

\begin{figure}
	\centering
	\includegraphics[width=\columnwidth]{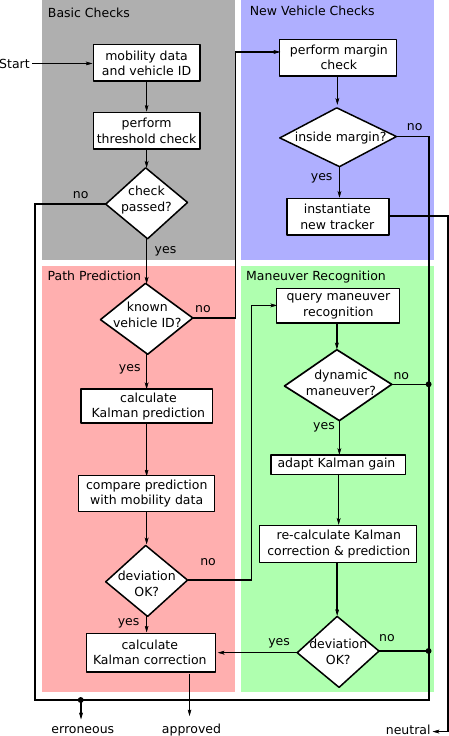}
	\caption{Figure based on \protect\cite{Stuebing2011-two-stage} explaining their detection method surrounding usage and updating of Kalman filters.}
	\label{fig:stuebing}
\end{figure}

The authors of~\cite{Stuebing2010-verifying} and~\cite{Jaeger2011-novel} verify transmitted \acp{CAM} by analyzing the sequence of messages to find the trajectory of each vehicle.
By tracking a vehicle using a Kalman filter, they can verify the location contained within each \ac{CAM}, thereby allowing the detection and correction of falsified data in \acp{CAM}.
This works, because the Kalman filter allows the accurate prediction of movement even under the influence of errors (e.g., they are also used to correct errors in GPS measurements).
As a result, the Kalman filter allows vehicles to locally link pseudonyms with high probability, and features adjustment for errors and new vehicles.
By defeating pseudonyms in this way, vehicles can check that vehicles are transmitting valid messages.
Their scheme explicitly does not distinguish between malicious and faulty nodes, instead aiming to detect any misbehavior.
We note that the existence of Kalman filters does not imply that privacy is void -- the Kalman filter only provides accurate estimates when actually following a vehicle (similar to physically following it by driving behind it).
The scheme was later extended~\cite{Stuebing2011-two-stage} to deal with situations where Kalman filter accuracy is poor, such as lane changes and other special maneuvers.
After computing the plausibility of a \ac{CAM} according to the Kalman filter associated with the sender, they propose a scheme that probabilistically recognizes the maneuvers of a vehicle using hidden Markov models.
A hidden Markov model is trained for each maneuver, classified by the Baum-Welch algorithm, allowing the state of the hidden Markov model to represent the different steps in the maneuver.
The information used for these models is then used to update the Kalman filters where necessary.
This entire process is shown in detail in Figure~\ref{fig:stuebing}.

\paragraph{Vehicle dynamics} Going beyond the position prediction work, which often assume a static vehicle model that is the same for every vehicle, Yavvari~et~al.~\cite{yavvari_vehicular_2017} suggest that prediction could be much more accurate.
Their core idea is to test the plausibility of new messages within the scope of old messages using an accurate model of the vehicles' dynamics.
This can then be used to construct bounding polygons that describe the extremes of valid positions that could be obtained given the vehicle dynamics and the previously received information.
Because this process requires detailed knowledge of the capabilities of each vehicle, the authors discuss how to derive this from BSM information.
In BSMs (in the US model), there is detailed information about the dimensions of each vehicle (with a resolution of 1cm).
This can be used to uniquely identify the vast majority of vehicle types, as shown by the authors.
Using this information allows for much tighter bounds than an abstract vehicle dynamics model.
Although this design raises some privacy concerns, the resources required to store or simulate the vehicle dynamics per vehicle type is relatively limited, and the bounds on valid behavior are quite tight.

\paragraph{Post-event detection} The idea of post-event detection is that events are not isolated from each other, but typically relate directly to future behavior.
A good example of such a scheme is the work by Ghosh~et~al.~\cite{Ghosh2010-detecting}, who looked at the post-crash notification (PCN) application.
After sending a PCN message, drivers normally adapt their behavior to avoid the crash site; this is exploited here to confirm whether the event was valid or not.
Although this specific scheme uniquely uses driver behavior for validation, other proposals have been made that use subsequent beacons to verify whether an event actually happened (e.g., Ruj~et~al.~\cite{Ruj2011} above).
In their scheme, Ghosh~et~al.~use a technique called root cause analysis to detect which part of the event message was false.
Although lacking in privacy, this scheme is very useful to serve as a baseline for revocation.
Post-event detection suffers if driver behavior models are unsuitable, and it notably cannot prevent false information from reaching the driver, which may lead to low user acceptance.
However, for future \acp{cITS} that rely on increased autonomous driving, post-event detection may be feasible, due to the fact that valid behavior is more well-defined.

\subsubsection{Consistency}
\label{sub:data_consistency}
Consistency-based mechanisms look at sequences of packets from distinct vehicles.
These mechanisms focus on detecting and resolving conflicting information to achieve an accurate representation of the real world scenario.
They are often employed by secure aggregation mechanisms to combine information from several vehicles into aggregates and to deal with inaccuracies, which may occur when aggregation mechanisms are used.

\paragraph{Raw data comparison} The simplest type of consistency mechanism is direct comparison between message contents to check them for conflicts.
This is closely related to consensus mechanisms from the node-centric mechanisms section: the mechanisms here are distinguished by the fact that a decision is mainly based on the data, rather than trust (potentially derived from data) in vehicles.
Unlike consensus mechanisms, direct checking mechanisms often simply output that a conflict exists (although some authors add some trust mechanism to verify the effectiveness of their scheme directly).

Leinm\"uller~et~al.~\cite{Leinmueller2006-improved,Leinmueller2008-decentralized} describe a position verification mechanism that bases on a number of different algorithms (also called sensors), each of which attempts to detect malicious or selfish behavior.
The position verification mechanism determines a trust value for each vehicle, but the focus of the paper lies on the sensors.
The authors propose sensors based on either consistency or plausibility (called cooperative and autonomous sensors respectively in the paper); we discuss the consistency sensors here, and the plausibility sensors in Section~\ref{sec:plausibility}.
The cooperative sensors are based on neighbor tables and position beacons to avoid the requirement of dedicated hardware.
First, pro-active exchange of neighbor tables can include positions or only include logical links between nodes.
In both cases, beacons are checked against received neighbor tables by comparing the claimed positions for a particular node in the beacon and the table.
When the tables do not include positions directly, nodes can extrapolate information using the maximum transmission range (using the accepted range threshold, see Section~\ref{sec:plausibility}).
Second, reactive position requests can be used as a more bandwidth-efficient sensor than periodic neighbor table exchange.
These requests are sent when an unknown vehicle $M$ is encountered; a vehicle knows the position of its neighbors, and selects a subset of them as either rejector or acceptor, based on whether the neighbor is in transmission range or not.
It then sends its request to this subset of neighbors, asking for the position of $M$.
Neighbors that do not know $M$ will respond with a corresponding message; others will respond with a position.
The sender can then compare the responses with the expected responses.
Both of these mechanisms rely on an honest majority, but are capable of dealing with noisy sensor data for those honest nodes.
The mechanisms can also cope with network loss, allowing the mechanisms to cope with limited attacks on lower layers.

\begin{figure}
	\includegraphics[width=\columnwidth]{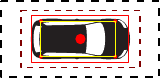}
    \caption{Figure based on \protect\cite{Bissmeyer2010} showing the different rectangles used to model the vehicle in the presence of GPS errors.}
	\label{fig:bissmeyer2010}
\end{figure}

In~\cite{Bissmeyer2010}, the authors use consistency between \ac{CAM} messages of vehicles to detect attacks.
The authors claim that a model typically used for crash avoidance systems can also be used to determine likely attacks.
The model describes that only one vehicle can occupy a given space at a given time.
This allows detection of falsified position information.
They assume it is possible to uniquely identify vehicles despite the use of pseudonyms (using the Kalman filter approach we discuss in Section~\ref{sub:plausibility}) and use this capability to determine whether vehicles intersect.
Intersection itself is measured using the given width and length of a vehicle in a \ac{CAM}.
A mechanism to deal with errors is also introduced, by simulating larger rectangles outside the bounds of the vehicle, as shown in Figure~\ref{fig:bissmeyer2010}.
The size of each of these is computed using the speed of the vehicle.
When detecting overlap of rectangles around different vehicles, a degree of certainty corresponding to the size of the rectangle is used.
The higher this certainty, the higher the probability that the position for one of two vehicles was falsified.
To compensate for lost packets, the authors also discuss a prediction mechanism to predict the location from which the next \ac{CAM} is sent.
To further increase detection accuracy, the authors note that a history of received \acp{CAM} and predictions can be used to compute a vehicles' trustworthiness; this trust value can then be used to decide which of two \acp{CAM} is the correct one.
This scheme is highly efficient in terms of bandwidth, but the accuracy of the detection may be limited due to the effects of noise in the sensors.
Although Sybil attacks are an issue for this work, the authors can deal with some degree of packet loss as well.

Rather than comparing individual messages between vehicles, Zaidi~et~al.~\cite{Zaidi2016} instead compare a single vehicles' data to the overall flow of all surrounding neighbors.
The authors propose to extend the beacon messages with three fields: flow, average speed and density, which all vehicles must compute and transmit regularly.
Flow is a traffic research term that is closely related to throughput; it is an indicator for congestion, and the flow computation used in this paper is related to speed and density.
The idea is that if an attacker falsifies its data, or uses Sybils to do so, there will be a significant difference between computed flow from benign and malicious vehicles.
Whenever detection occurs, a vehicle is monitored for some time; if the vehicle is marked as malicious (through hypothesis testing), the vehicles data is rejected.
This scheme has the significant advantage that the proof of misbehavior is rigorous (due to the hypothesis testing approach), but suffers from attacks when the attacker slowly increases their value (and those values of controlled Sybils).
This attack is excluded by the authors attacker model.
Also note that this scheme only works on a per-road basis (messages between different roads will not verify, since they have different flow); this could be resolved by including road identifiers.

Grover~et~al.~\cite{Grover2011-sybil-attacks} perform Sybil attack detection by comparing neighbor tables of several vehicles over time.
The reasoning behind their work is that the fake identities of the attacker must always be in the vicinity of the attacker in order to be able to the necessary local majority the attacker wants to achieve.
Therefore, they must frequently appear in the neighbor tables of legitimate vehicles as a group, while legitimate vehicles will not form such a group over time.
Because network topology is highly dynamic in \ac{cITS}, the authors distinguish between uni-directional and bi-directional communication links.
We note that while the authors present an effective way to limit Sybil attacks in duration, the communication overhead is significant, as is the latency for detection.
In addition, certain traffic scenarios cause the assumption of recurrence to become unrealistic, e.g., in traffic jams, which increases false positives or increases detection latency.
Most importantly, it does not protect against Sybil attacks that have a short duration, which however may be sufficient to attack a variety of information dissemination mechanisms, such as traffic jam detectors.

\cite{Xiao2006-detection} describes one of the earliest of such mechanisms, verifying claimed positions using signal strength metrics.
The basic scheme consists of three roles: claimer, where a node claims a position in a beacon, witness, where a node receives a beacon and measures its proximity using the received signal strength, which it transmits in subsequent beacons, and verifier, which collects signal strength measurements to estimate and verify the position of a node.
To ensure that the scheme is not subject to Sybil attacks, the authors design an improved position verification scheme, which focus on the witnesses.
In this scheme, \acp{RSU} issue a signature of proximity at a specific time, with a defined driving direction.
By authenticating direction, the authors can reliably establish which witnesses are approaching vehicles, and only use these for detection.
The scheme does not account for scenarios with low amount of witnesses, or one-way situations, but their statistical approach that compares RSSI-based estimations with claimed positions is theoretically sound.
After classifying a neighbor as a Sybil node, the vehicle will go through its neighbor table and verify other neighbors in order to find other Sybils originating from the same malicious entity.
A final criticism is that this scheme assumes Sybil attacks are targeted; attacks aiming to confuse approaching vehicles by issuing false witnesses continuously are still possible.

\paragraph{World modeling} Rather than comparing raw data, some authors suggest the more abstract approach of world modeling.
This removes the restriction of comparison between specific messages, and rather allows queries such as ``is this world state consistent?'' or ``is this hypothesis true''?
For detection, these approaches often aim to find the data that is the cause of a conflict in the consistency of the model.

Golle~et~al.~\cite{Golle2004} have presented the earliest example of data-centric detection, which checks for consistency between assertions (messages, also called observed events in the paper) using some abstract model of the \ac{cITS}.
If this results in an invalid state, their system identifies possible explanations where a subset of all assertions is valid -- the best explanation can then be chosen.
The model relies on four core assumptions: nodes can bind observations to received communication, they can uniquely identify neighboring vehicles (that is, detect Sybil attacks using physical properties), they can authenticate to one another, and finally, the network graph should always be connected.
In case inconsistencies are found, Occam's Razor is applied, meaning that the explanation with the least amount of attackers best explains the conflicts found.
As an example of how their model works, the authors introduce two models that determine the correct locations of vehicles in the network.
Although this is one of the earliest works on data-centric security mechanisms in \acp{cITS}, the outline of their mechanism is still used as a guideline for data-centric security and secure aggregation schemes.
Unfortunately, the detection component of this work is not evaluated for feasibility, due to an assumption that neighbors can immediately exchange derived information.
As the authors note, assuming this instant connectivity is not very realistic, as the amount of data exchanged in their scheme is quite high and available bandwidth in \acp{cITS} is limited.
In addition, it seems to be assumed that the attacker is a protocol-adhering participant, e.g., there are no attacks on routing or on lower communication layers (such as selective jamming or altered transmission power).

Data-centric security mechanisms have also been applied to secure aggregation, as proposed by~\cite{Dietzel2010}.
Aggregation is here motivated as a means to make \acp{cITS} more feasible.
Due to high potential bandwidth requirements, especially for traffic efficiency applications that require information from large areas, aggregation will be essential to efficiently disseminating information through the \ac{cITS} on a large scale.
However, security is an unresolved issue in aggregation: in particular, aggregation aims to remove redundancy of information, which increases the challenge of applying data-centric security mechanisms.
The authors specify a framework using fuzzy reasoning to detect that an attack is in progress with high accuracy.
As input for this reasoning process, clues attached to each aggregate are used, which are atomic, unaggregated observations signed by their original senders.
These clues provide the necessary integrity that is typically hard to obtain in secure aggregation.
Using fuzzy logic to determine confidence in data, rather than trust in nodes, is a main advantage of the scheme.
The attacker model in this paper considers an attacker that is specifically aimed at compromising aggregation mechanisms, rather than a general attacker that also exploits lower layer and Sybil attacks.

In~\cite{Bissmeyer2012}, the authors propose a more strongly centralized approach.
They require each node to detect incidents, and forward them to a central authority that will detect misbehavior based on these reports.
Because the central authority only needs to perform detection based on reports, rather than continuously for all nodes, this provides a scalable approach.
In addition, this makes the assumption of an honest majority reasonable, when combined with revocation.
Each node assigns trust to all nodes it knows; confidence in this trust is defined as a weight on the trust assigned by another node.
For example, if node $a$ has neighbors $b, c$ and assigns a high trust to $b$, while $b$ assigns a low trust to $c$, then confidence is used to describe the amount of evidence each node has for its trust assignment.
These parameters are later used to determine reputation of each node by the central authority.
The central authority can then resolve pseudonyms and determine whether any node was cheating repeatedly, or whether there was a benign fault.
This is done by using the trust and confidence of the suspect and reported nodes; attackers are detected by determining which nodes have very low reputation.

\paragraph{Rule mining} Finally, some authors have proposed mechanisms that use techniques from the field of data mining to extract useful information from a large amount of stored data, such as \acp{CAM} and \acp{DENM}.
By extracting these rules, the likelihood of the message being valid can be expressed.

VARM~\cite{Rezgui2011-detecting} is an example of this idea, which directly applies data mining techniques to misbehavior detection.
The authors propose a data mining-based mechanism that dynamically derives association rules from received data using a data-structure called the Itemset-Tree.
Association rules express correlations in a data set, and are a basic concept in data mining.
By extracting these association rules, the node thus infers information from received messages that represent the expected behavior of senders.
This knowledge can express hidden information that represents local road conditions, without the need to list all such scenarios and develop rules or models for them by hand.
Their core drawback is that they may not generalize, because correlation does not imply causation.
Another issue is that the paper does not extensively study the bandwidth requirements posed by their scheme, nor is latency or detection rate the main target of the study.
We note that data mining is typically applied in scenarios where latency and computational resources are not an issue, and these techniques may not provide sufficient performance.
Nevertheless, the application of data mining is a novel idea that can combine elements from both data-centric and behavioral misbehavior detection; this work provides a good starting point for applying data mining techniques to misbehavior detection.
VARM is considered to be a consistency-based misbehavior detection mechanism because the authors specifically focus on temporal relationships between events received from many different vehicles, rather than verifying individual vehicles.

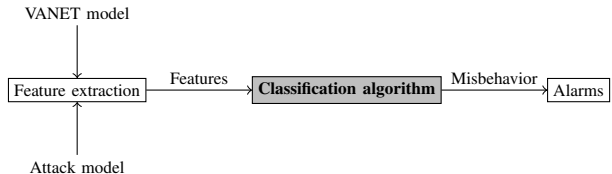
\begin{figure}
\centering
  \begin{tikzpicture}[scale=0.65, every node/.style={scale=0.65}]
\node (vanet) {VANET model};
\node[below=2em of vanet,draw] (feature) {Feature extraction};
\node[below=2em of feature] (attack) {Attack model};
\draw[->] (vanet) edge (feature)
          (attack) edge (feature);
\node[right=4em of feature,draw,fill=lightgray,font=\bfseries] (ca) {Classification algorithm};
\node[right=4em of ca,draw] (alarms) {Alarms};
\draw[->] (feature) edge node[above] {Features} (ca);
\draw[->] (ca) edge node[above] {Misbehavior} (alarms);
\end{tikzpicture}
\caption{Machine learning architecture proposed in \protect\cite{Grover2011-machine-learning}.}\label{fig:machine-learning}
\end{figure}

\paragraph{Machine learning}
Grover~et~al.~\cite{Grover2011-machine-learning} have also proposed a machine learning-based method to detect misbehaving vehicles.
Grover~et~al. have implemented several attacks on routing and several attacks on message content, in particular a Sybil attack and position forging.
They use machine learning to learn the features of legitimate and misbehaving messages, based on models of both.
The authors defined several features and used a network simulator to generate traffic; the features of this traffic was used to train several classification algorithms.
There are two classes of features -- one used for detecting position falsification and Sybil attacks (similar to many data-centric mechanisms) and one class that considers various temporal aspects and delivery ratios.
The generated network traffic was created using a two-way, multi-lane highway scenario.
The authors used the values for each of these features as input for several classification algorithms from machine learning, contained in the WEKA toolset~\cite{weka}.
The best results were obtained using the Random Forest~\cite{random_forest} and J48 algorithms.
Figure~\ref{fig:machine-learning} shows the overall architecture of the proposed mechanism.
In a later paper~\cite{Grover2012}, they improved on their previous work by replacing the single classification algorithm with a number of classification algorithms.
After classification by each of the algorithms, a majority decision is used to decide whether a given situation is considered part of an attack.
In both papers, specific implementations of attacks are used to evaluate the system; these implementations are used to generate data for both the training and testing set.
Both of these works rely on specific implementations of attacks and a specific scenario; in particular, no details are provided regarding attack implementation and the base scenario.
It is thus difficult to determine whether these classifiers provide a general solution, or a solution specific to the scenario, since it is unknown how real-world behavior compares to the simulated legitimate behavior.
 
Ghaleb~et~al.~\cite{Ghaleb2017} propose a similar approach to Grover~et~al., using neural networks.
The authors describe how vehicles collect information from their environment using sensors, which is then shared through communication (i.e., through \ac{V2V} communication).
This information is then used to construct a local dynamic map (LDM) of a vehicle, from which features for misbehavior detection can be derived.
Historical values of these features are used to construct a model, which is then used in an on-line fashion to classify new messages into legitimate and malicious.
In their work, features include a variety of plausibility and consistency checks that they cite from previous work, as well as some own features.
The authors use a simulated set of misbehavior to create a supervised learning problem, to which they apply their neural network, which is a standard feed-forward back-propagation network with 7 input neurons and 15 neurons in one hidden layer.
The output layer is not described, but can be assumed to have a single neuron (i.e., misbehavior or not).
The evaluation of this work has significant weaknesses, including a communication model far behind the state of the art and no details on the implemented attacks.
The authors also use the entire dataset for training, before using four vehicles from this dataset for testing (which is methodologically unsound).
Results were compared to earlier work, whose results correspond to some of the features proposed in the feature collection (it is not clearly stated which of the work they compare to, as they reference multiple papers).
Collecting the data necessary to build a model for wide-spread, static deployment requires a lot of resources, and the resulting model could easily be used by an attacker to search for messages that will be accepted by a receiver.
However, the approach described by the authors is easy to generalize to other areas where appropriate data and features are available.

The authors in~\cite{Yan2011-general} focus on discussing the verification of location data by using additional vehicle sensors that allow detection by line-of-sight, so-called eye-devices, in order to verify received messages (which are referred to as coming from ear-devices).
The authors propose that each vehicle should have a variety of radars and cameras to detect other vehicles.
It is assumed that $85\%$ of all vehicles is honest, and that \ac{GPS} data can be trusted if the information corresponds to the own location combined with the data from eye-devices.
However, eye-devices only work for line-of-sight; thus, the authors allow vehicles to request eye-device confirmation from vehicles on the oncoming traffic.
Similarly, neighbors can be queried, although oncoming traffic is preferred.
The authors express the eye data through vectors relative to the current vehicle (for speed and location).
The authors then go on to define local security as verification by checking against local sensors, including ear-devices (messages from different vehicles).
The authors claim that local security can be used to obtain global security, reasoning that this is analogous to greedy algorithms.
However, we note that greedy algorithms are heuristics, which by definition do not necessarily provide a global solution.
As the paper lacks any further proof of a globally optimal solution, it is likely that their mechanism can achieve at best a local solution.
In addition, the authors do not seem to consider errors in the obtained locations and other sensor data, further weakening their claim.
Finally, in low density scenarios the attacker may be able to exploit the mechanism to cause false detections using specially crafted messages, although this is a fairly obscure scenario.


%
\newcommand\subhead[1]{%
	\addlinespace
	\multicolumn{9}{l}{\emph{#1}}\\
	\midrule%
}

\newcommand\tbnl{\\\addlinespace}

\newcommand\positive{\tikz{%
	\draw[thick] (0,0) circle (1ex);
	\draw[very thick, scale=.7] (-1ex,0) -- (1ex,0) (0,-1ex) -- (0,1ex);
}}
\newcommand\neutral{\tikz{%
	\draw[thick] (0,0) circle (1ex);
}}
\newcommand\negative{\tikz{%
	\draw[thick] (0,0) circle (1ex);
	\draw[very thick, scale=.5] (-1ex,0) -- (1ex,0);
}}
\subsection{Overview}\label{sec:tableoverview}
Now that we have provided an overview of different types of mechanisms that exist in the literature, we summarize these mechanisms and their properties in Tables~\ref{tab:mechAssess1} and \ref{tab:mechAssess2}.
These tables contain the most important pros and cons for each mechanism, as well as a qualitative analysis of the scope of detection, required resources (which includes computational and communication resources), generalizability, security and impact on privacy.
The scope of detection tells us whether a mechanism performs local (L), cooperative (C) or back-end (B) detection, connecting back to the discussion on detection scopes in Section~\ref{sec:scope}.
The required resources field in the table gives an intuition of how much bandwidth, computational resources and memory is required; highly efficient mechanisms are marked with \positive, while expensive mechanisms are marked with \negative.
Good generalizability (\positive) means that the mechanism may be transferred and applied to other domains relatively easily, a topic we discuss in detail in Section~\ref{sub:applications_beyond_cits}.
With the security column, we illustrate how easily a mechanism can be manipulated: a mechanism subject to certain attacks is marked with \negative, while a mechanism with a limited attacker model (compared to the model discussed in Section~\ref{sec:attacker-model}) is marked with \neutral{} and a good scheme is marked with \positive.
Note that this security column does not take into consideration the detection accuracy, but only the security of the detection mechanism.
The last column, privacy, indicates the necessary type of linkability, as discussed in Section~\ref{sec:pseudonyms}; full linkability (F), explicit linkability (E), implicit linkability (I) or no linkability (N).
When the issue is not addressed in the paper, we use a question mark.
\begin{table*}
  \centering
	\caption{Mechanisms overview\label{tab:mechAssess1}}
  {
\begin{tabular}{L{2.6cm}p{4.5cm}p{4.5cm}cccccc}
	\toprule
	Mechanism & Pros & Cons &
	\parbox[b][][t]{1em}{\rotatebox{270}{Scope}} &
	\parbox[b][][t]{1em}{\rotatebox{270}{Resources}} &
	\parbox[b][][t]{1em}{\rotatebox{270}{Generalizability}} &
	\parbox[b][][t]{1em}{\rotatebox{270}{Security}} &
	\parbox[b][][t]{1em}{\rotatebox{270}{Privacy}} &
	$\sum$ \\
	\midrule
\subhead{Node-centric / behavioral (Section~\ref{sub:behavioral_mechanisms})}


Watchdogs~\cite{Hortelano2010} &  Low implementation complexity  &  Only detects malevolent forwarding, vulnerable to attacks & L & \positive & \positive & \negative & ? & \negative \tbnl
\midrule
Jamming detection~\cite{Hamieh2009,Punal2012-VANETs} & Detects adaptive jamming & No attacker identification & L & \positive & \positive & \positive & N & \positive \tbnl

\subhead{Node-centric / trust-based (Section~\ref{sub:node_trust_based})}

Urban path-based~\cite{Chen2009-robust} & fast, computationally efficient detection & infrastructure-dependent, privacy issues, vulnerabilities and false positives & C & \negative & \negative & \negative & ? & \negative \tbnl

P$^2$DAP~\cite{Zhou2011-p2dap} & Strong detection/attacker model; potentially efficient approach & Infrastructure dependent; no privacy to central authority & B & \positive & \neutral & \positive & E & \neutral \tbnl

Footprint~\cite{Chang2012} & Reasonable level of privacy without high cost; useful for authenticating events &  infrastructure-dependent; geared only towards Urban networks; full trust in \acp{RSU}; potential false positives & C & \neutral & \negative & \negative & E,I & \negative \tbnl

\midrule

LEAVE~\cite{Raya2007-Eviction} &  Aggregation improves the quality of detection and reduces latency  &  Mechanism is prone to Sybil attacks & C & \neutral & \positive & \neutral & ? & \neutral \tbnl

OREN~\cite{Bilogrevic2011,Bilogrevic2010}~\cite{Raya2010} & Game-theoretic evidence, integrates issuance of pseudonyms as incentive system, pseudonyms are not linkable & Prone to Sybil attacks by financially capable attackers & B & \positive & \neutral & \neutral & N & \negative \tbnl

 SLEP~\&~PRP~\cite{Zhuo2009} & Holistic approach, integrating cooperative and back-end detection & Potentially prone to Sybil attacks & B & \neutral & \neutral & \neutral & E & \neutral \tbnl

  Sowattana~et~al.~\cite{sowattana_distributed_2017} & Improves ideas from earlier work & Voting is prone to Sybil attacks, expensive neighbor table exchange & C & \negative & \negative & \negative & E & \negative \tbnl
\midrule

Decision logics~\cite{Raya2008} & Can reuse results of the mechanisms & No optimal decision logic & L & \positive & \positive & \neutral & E & \neutral \tbnl

Bayesian logic-based~\cite{Rawat2011-securing} &  Suspiciousness instead of trust & Misclassification of unusual events & L & \negative & \positive & \positive & E & \negative \tbnl

\midrule

PoR~\cite{Cao2008,Kamra2006-mag} & Simple and relatively efficient & Decoding issues for new nodes, weak attacker model & C,B & \positive & \positive & \negative & F & \negative \tbnl

Z-smallest probabilistic counting~\cite{Hsiao2011} &  Protection against inflation attacks, bandwidth efficient broadcast  & Enforces a specific message format, weak attacker model  & C & \neutral & \positive & \neutral & E,I & \neutral \tbnl

Cooperative trust building~\cite{Leinmueller2010-its-congress} &  Builds on simple schemes, trust transitivity for improved performance &  Bandwidth requirements not considered  & C & \neutral & \positive & \neutral & ? & \positive \tbnl


Dynamic threshold trustworthiness~\cite{Petit2011} &  Maximizes the confidence in messages until response is required  &  Potentially vulnerable to targeted attacks exploiting the mechanism  & C & \positive & \positive & \negative & ? & \neutral \tbnl

Certainty of Event~\cite{Kim2010} & Intuitive representation, combines security and safety aspects & Highly application-specific, relies on verified positions & C & \positive & \negative & \neutral & E & \positive \tbnl



	\bottomrule
\end{tabular}
}
\end{table*}

\begin{table*}
  \centering
  \caption{Mechanisms overview (continued)\label{tab:mechAssess2}}
{
\begin{tabular}{L{2.6cm}p{4.7cm}p{4.7cm}cccccc}
	\toprule
	Mechanism & Pros & Cons &
	\parbox[b][][t]{1em}{\rotatebox{270}{Scope}} &
	\parbox[b][][t]{1em}{\rotatebox{270}{Resources}} &
	\parbox[b][][t]{1em}{\rotatebox{270}{Generalizability}} &
	\parbox[b][][t]{1em}{\rotatebox{270}{Security}} &
	\parbox[b][][t]{1em}{\rotatebox{270}{Privacy}} &
	$\sum$ \\
	\midrule
\subhead{Data-centric / plausibility (Section~\ref{sub:plausibility})}

Signal Analysis~\cite{Guette2007-sybil} & Thorough analysis of RSSI-based estimation & Potentially infeasible with GPS errors & L & \positive & \negative & \neutral & N & \positive \tbnl

Transmission time-based~\cite{Ruj2011} & Useful for specific events &  Unrealistic assumptions for position verification, vulnerable to attacks  & L,C & \positive & \negative & \negative & E & \negative \tbnl

  EKF with AoA and DS~\cite{Sun2017} & Efficient tracking of malicious vehicles & Additional hardware required, relies on privacy breach & L & \positive & \negative & \positive & E & \positive \tbnl

  RSSI voiceprint~\cite{yao_multi-channel_2018} & Linking Sybil nodes through physical characteristics & Attacker may only use one radio & L & \positive & \positive & \neutral & I & \positive \tbnl
\midrule

  \cite{Leinmueller2008-decentralized}, \cite{Leinmueller2006-improved}~and~\cite{vanderHeijden2016} &  Efficient  &  Limited detection capabilities & L & \positive & \negative & \negative & I/N & \neutral \tbnl

PVN~\cite{Lo2007} &  Fast and efficient, allows relations between plausibility checks  &  Insufficient as the only means to detect attacks  & L & \positive & \negative & \neutral & I & \positive \tbnl

VEBAS~\cite{Schmidt2008} &  Self-sufficiency of mechanisms  &  Honest majority assumption  & L,C & \positive & \neutral & \positive & F & \positive \tbnl


\midrule

Kalman filter-based~\cite{Stuebing2010-verifying,Stuebing2011-two-stage,Jaeger2011-novel} &  Highly efficient tracking of vehicles &  Computationally expensive, relies on partial privacy breach & L & \positive & \negative & \positive & I & \positive \tbnl

\midrule

  Prediction based on vehicle dynamics~\cite{yavvari_vehicular_2017} & Excellent performance & Ease of tracking and potentillay large database & L & \neutral & \negative & \positive & E & \positive \tbnl

\midrule

After-the-fact verification~\cite{Ghosh2010-detecting} &  Highly useful for reporting  & No preemptive detection, each application requires accurate predictions & L & \positive & \neutral & \positive & ? & \positive \tbnl

	\bottomrule
\end{tabular}
}
\end{table*}

\begin{table*}
  \centering
  \caption{Mechanisms overview (continued)\label{tab:mechAssess3}}
{
\begin{tabular}{L{2.6cm}p{4.7cm}p{4.7cm}cccccc}
	\toprule
	Mechanism & Pros & Cons &
	\parbox[b][][t]{1em}{\rotatebox{270}{Scope}} &
	\parbox[b][][t]{1em}{\rotatebox{270}{Resources}} &
	\parbox[b][][t]{1em}{\rotatebox{270}{Generalizability}} &
	\parbox[b][][t]{1em}{\rotatebox{270}{Security}} &
	\parbox[b][][t]{1em}{\rotatebox{270}{Privacy}} &
	$\sum$ \\
	\midrule

\subhead{Data-centric / consistency (Section~\ref{sub:data_consistency})}

Overlap detection~\cite{Bissmeyer2010} & Good detection results under reasonable assumptions & Requires fixed transmission power & C & \positive & \negative & \positive & I & \positive \tbnl

Abnormal flow detection~\cite{Zaidi2016} & Novel idea that applies traffic research knowledge & Constrained to specific roads, vulnerable to gradual change & L & \neutral & \negative & \neutral & N & \neutral \tbnl

Neighbor-sets over time~\cite{Grover2011-sybil-attacks} & Novel ideas, prevents long-term Sybil attacks & False positives possible; no prevention for short-term Sybil attacks & C & \negative & \neutral & \neutral & ? & \neutral \tbnl

Enhanced Position Verification~\cite{Xiao2006-detection} &  Works well in ideal situations  &  Depends on infrastructure \& network properties, vulnerabilities & L(,C) & \positive & \negative & \negative & ? & \neutral \tbnl

Proactive cooperative sensors in~\cite{Leinmueller2008-decentralized} & Simplicity, capable of dealing with noisy data, powerful in dense scenarios & Vulnerable to Sybil attacks, high bandwidth cost & C & \negative & \neutral & \neutral & ? & \neutral \tbnl

Reactive cooperative sensors in~\cite{Leinmueller2008-decentralized} & Simplicity, capable of dealing with noisy data & Unreliable detection, requires unicast message exchange & C & \neutral & \neutral & \positive & ? & \positive \tbnl

\midrule

Attacker modeling~\cite{Golle2004} & Structured and general approach & Requires complex reasoning techniques on every vehicle & C & \negative & \neutral & \neutral & I & \neutral \tbnl

Resilient secure aggregation~\cite{Dietzel2010} &  Efficient approach to secure aggregation  &  only works for aggregation purposes  & C & \neutral & \negative & \positive & ? & \positive \tbnl

Back-end detection~\cite{Bissmeyer2012} &  Uses global knowledge to improve/confirm detection  &  Potential privacy concerns; potential high cost at the back-end & B & \negative & \positive & \positive & E & \positive \tbnl

\midrule

VARM~\cite{Rezgui2011-detecting} &  Novel approach, based on well-researched area  &  	Feasibility and mapping of rules to conclusions is unclear  & L & \negative & \positive & \neutral & ? & \positive \tbnl

\midrule

Machine-learning~\cite{Grover2011-machine-learning}~\cite{Grover2012} &  Machine learning allows easy generalization & Prone to gradual changing from attacks, decisions not transparent & L & \neutral & \positive & \neutral & ? & \neutral \tbnl

Neural Networks~\cite{Ghaleb2017} & Generalizability & Fixed model, weak evaluation, privacy unclear & L,B &\negative &\positive &\negative &? &\negative \tbnl

\midrule

Active position detection~\cite{Yan2011-general} & Dedicated hardware for secure positioning is highly effective & Hardware requirements, no privacy, vulnerable to Sybil attacks & C & \positive & \negative & \negative & ? & \negative \tbnl

	\bottomrule
\end{tabular}
}
\end{table*}

\section{Solved and Open Challenges}
\label{sec:solved_and_open_challenges}

Starting with an overview of the current \ac{cITS} ecosystem, as well as ongoing standardization efforts, we have provided an extensive categorization of different misbehavior detection mechanisms.
In this section we discuss the solved and open challenges for security in \ac{cITS}.
In addition, we provide a discussion of ways to transfer the security mechanisms to other fields.

\subsection{Reoccurring patterns}
\label{sec:reoccurring_patterns}

In our discussion of the mechanisms, we often pointed out potential attacks tied to other challenges, which are solved by different mechanisms.
Indeed, many mechanisms in our discussion are complementary, and here we discuss several recurring patterns of approaches taken by these mechanisms.

\paragraph{Physical models} A number of mechanisms use physical models to detect misbehavior~\cite{Leinmueller2008-decentralized,Bissmeyer2010}.
These models allow a detection mechanism to incorporate specific knowledge about the driving process of vehicles.
The complexity of models ranges from very simple to very complex models.
An example of a simple model is that cars cannot drive faster than 500\,km/h, or that two cars cannot occupy the exact same position.
Complex models include an analysis of acceleration and deceleration behavior, movement prediction including turn probabilities, correlation of vehicle positions to street maps and predictions based on vehicle dynamics~\cite{Sun2017}.
Sensors are often used as a base truth by the observing vehicle; it remains to be investigated whether attacks can be designed that affect sensors and messages simultaneously in a consistent way.
Regardless, the correlation with models is a very interesting approach, because it often works locally and does not depend on correlation of data from several vehicles (which is also a common attack vector).
Therefore, detection components based on physical models should be included in a misbehavior detection system.

\paragraph{Signal properties} Besides physical models targeted at the process of driving itself, physical properties of the wireless channel (i.e., signal properties) are another important tool for detection of spurious messages.
The typical wireless reception range can be estimated, and moreover, WiFi hardware is able to measure the strength of a received signal.
When vehicles synchronize their clocks with GPS, time of flight measurements can also be used.
All these observations can be used to detect messages which are, for instance, received with a high signal strength but claim to originate from a position far away.
In addition, there have been proposals to include additional hardware (typically additional antennae, see~\cite{Ren2009-location}) to effectively perform more advanced signal analysis to determine the direction from which the signal originates.
However, the effectiveness of signal-based mechanisms is somewhat disputed due to the high mobility, cost constraints and potential inaccuracy~\cite{Leinmueller2006-improved,Guette2007-sybil}.
This will be especially challenging when combined with newly proposed MAC-protocols that exploit the additional spatial re-use enabled by transmit power variations, such as decentralized congestion control in ETSI~\cite{etsi:ts102687}.
Although some proposals~\cite{yao_multi-channel_2018} address this issue by looking at signals over time, this increases the detection latency.

\paragraph{Machine learning techniques} A problem that many mechanisms face is how to interpret locally collected data in order to find patterns, or how to merge results from a set of different misbehavior detectors into one coherent output.
To solve these problems, a number of schemes adapt ideas from machine learning.
Often-employed approaches include decision trees, Bayesian inference or Dempster-Shafer theory~\cite{Raya2008,Rawat2011-securing,Rezgui2011-detecting}.
These approaches can provide useful tools to analyze data and derive certain interesting features that might point at attacks.
Some reputation systems, such as those proposed in~\cite{Bilogrevic2010}, already use some of these ideas to manage node reputation, but these typically only focus on node-centric aspects, such as how to maintain the reputation system and how to protect it against attacks.
In addition to providing misbehavior detection for vehicles, these node-centric approaches may be later used to perform an after-the-fact analysis, which in turn can be used to verify detected misbehavior.
This is essentially the idea behind various data-centric machine learning approaches, where a large volume of data is labeled and used to learn a classifier.
For this latter case, collecting labeled data that is sufficiently diverse is a huge challenge.

\paragraph{Centralized detection} Early works often perform the whole process of misbehavior detection locally and only send reports about local decisions to the back-end to perform pseudonym resolution and revocation.
In these works, the back-end does not perform any detection task; rather, it just verifies received reports for correctness and checks for false accusations.
However, performing detection exclusively in a local fashion prevents the detection mechanism from detecting larger clusters of distributed misbehavior.
For instance, an attacker could repeatedly mount attacks at different locations, which might be detected locally, but only in correlation show the full extend of the attacks' severity.
Hence, newer schemes include the back-end more and more in the actual detection work.
Notably~\cite{Bissmeyer2012} and~\cite{Zhou2011-p2dap} employ such centralized detection approaches.
A challenge to solve in this context is the level of event reporting.
In order to allow the back-end to detect more attacks than possible locally, it may be necessary to report more data to the back-end, including even mildly suspicious behavior, to provide a large enough data basis for centralized detection.

\subsection{Open Issues}
\label{sec:open_issues}

Even though many specific aspects of misbehavior have been addressed, there are still a number of open issues.
We discuss these issues based on our survey of the state of the art and the mechanisms discussed in Section~\ref{sec:state_of_the_art}.
These issues are challenges regarding the detection itself -- dealing with conflicting data, for instance by discarding or hiding messages, error correction, reporting and so forth is an orthogonal issue.

\paragraph{Thresholds for detection}
\label{par:thresholds}

An important problem in any intrusion detection system is its configuration; after which threshold is a message malicious? This question is also an important challenge in misbehavior detection, because high false positive or false negative rates can cause significant problems or even cause the system to be rejected entirely.
Especially for \acp{cITS}, where direct reporting to an expert through warnings is not always possible, this needs to be addressed.
As a related issue, the distinction between erroneous and explicitly malicious messages would be useful, although this may not be feasible.
Additionally, because of the data-centric nature of many detection mechanisms, the question of when information should be merged is significant, and this may affect detection performance directly.

\paragraph{Identification of misbehaving nodes}
\label{par:identification_of_misbehaving_nodes}

We note that while data-centric mechanisms are superior at detecting conflicting data and identifying which data is malicious, they may not be able to identify the attacker directly.
Although for many applications, it is more important to have accurate data than it is to identify attackers to mitigate possible harm through attacks, node-centric mechanisms that identify attackers are still useful.
They can be used to perform local or global revocation, thereby limiting both the amount of attackers in the network and the impact they may have on the network.
As an example take the fake traffic jam scenario from Figure~\ref{fig:fake-message}.
Although a data-centric mechanism may detect that the reported traffic jam is fake, without proper identification of the sender, a local revocation is not feasible.
Finally, identifying which data belongs to which vehicle may provide additional input for node-centric mechanisms, for instance to assign a credibility to information sources.

\paragraph{Voting, pseudonyms and Sybil attacks}
\label{par:voting_using_pseudonyms}

Many node-centric mechanisms rely on long-term identities to perform voting or other trust-based evaluation of data.
The basic idea is simple: assuming that an event, such as an icy road, is detected by all passing vehicles and further assuming that the majority of vehicles is honest, a simple majority vote about whether the road is actually icy should reveal possible misbehavior.

However, it is a common assumption that vehicles will use short-term pseudonyms for communication to protect privacy, as discussed in Section~\ref{sub:its-privacy}.
Therefore, an important open issue is to build a scheme that performs voting using pseudonyms, while still providing the necessary resilience against Sybil attacks.
A simple solution would be to require that a vehicle can only use one pseudonym at a time, but this assumption is disputable both from a practical and a privacy point of view.
Although some work has already been done (see Section~\ref{sub:node_trust_based}), it is an open challenge to design a voting scheme where all participants can use arbitrary pseudonyms while still allowing to reveal multiple votes by the same vehicle.
One possible solution for this challenge is to assume limited processing capabilities for the attacker and employ a cryptographic primitive called \acf{PoW}.
As the name implies, this primitive allows a vehicle to prove that it has performed some amount of computation, and it is traditionally employed to hinder denial of service attacks.
This works because the computation becomes exponentially harder and cannot be precomputed.
One proposal to use \ac{PoW} mechanisms for \acp{cITS} is~\cite{Palomar2012-PoW}, which requires \ac{PoW} for evidence that an event actually occurred.
The problem with these approaches is that they pose significant computational overhead, and thus cannot work in all situations.
An alternative would be assuming a functioning revocation infrastructure with strong penalties.
This significantly reduces the prevalence of malicious nodes, since they would be removed from the network, which reduces the required false negative rate.
However, this does not protect the \ac{cITS} against stronger, irrational attacker models, since these will not care about penalties.

\paragraph{Linkable messages}
\label{par:resolving_pseudonyms}

While the aforementioned mechanisms are beneficial for detecting misbehavior, they might have negative impact on the location privacy the users.
As discussed above, pseudonym schemes are employed to prevent attackers from collecting location traces of specific vehicles over a longer time.
The standardization efforts aim mainly at preventing privacy impact on a larger scale: the goal is to prevent attackers from efficiently following many participants, given a few stationary attacker-controlled network nodes.
This is opposed to attackers that track a single vehicle with a moving node -- this is, after all, equivalent to physically following a vehicle.
It has been shown by privacy research~\cite{Wiedersheim2010} that it is possible to track vehicles in this manner, which is indeed necessary for certain safety applications, such as those that rely on path prediction.
This functionality is provided by employing movement models for vehicles to predict future positions and correlate them with current positions.
Such linking can provide the necessary information, which can not only enable \ac{cITS} applications, but also misbehavior detection.
Indeed, several data-centric mechanisms, such as~\cite{Stuebing2011-two-stage}, already exploit this approach.
However, it is an open challenge to assess the exact privacy impact of these mechanisms and their full potential for misbehavior detection.
In particular, the reliability of these mechanisms can subsequently have a large impact on accuracy, which an attacker could exploit.
We remark that mechanisms relying on these approaches may be defeated by future, more powerful privacy mechanisms.

\paragraph{Cooperative misbehavior detection}
\label{par:cooperative_misbehavior_detection}

As long as vehicles or other entities only analyze locally received data to detect misbehavior, trust in the derived opinions is not a big issue.
That is, we can assume that vehicles trust the mechanisms running locally.
Even though the decisions reached might be associated with uncertainty, that uncertainty is still known.
Once vehicles exchange reports with other vehicles, the situation is different.
A misbehavior report received by another vehicle is basically just another item of information, like any other information item received from remote vehicles.
Therefore, we have to assume that the misbehavior reports can be fake as well.
Some schemes propose to solve this issue by collecting a number of misbehavior reports from different vehicles, again using a voting mechanism, which relies on long-term identities.
Other approaches propose to solve the issue by so-called suicide schemes where vehicles reporting misbehavior of other vehicles also exempt themselves from the network (see Section~\ref{sub:node_trust_based}).
However, another large share of related work ignores the issue of trusting cooperative misbehavior reports.
These schemes assume that reports from other vehicles are trustworthy if they carry correct signatures.
An ideal solution for cooperative misbehavior detection has not been presented yet.

\paragraph{Level of reporting to back-end}
\label{par:level_of_reporting_to_backend}

As discussed above, we need to rethink the level of reporting to the back-end in order to benefit from misbehavior detection in the back-end.
If the back-end only receives reports about definite misbehavior, then local nodes have already identified the misbehavior.
Hence, the additional benefit of involving the back-end is limited to possible revocation of the misbehaving nodes' certificates.
On the other hand, it is not possible to report all data received by vehicles to the back-end because of bandwidth constraints.
It is an open challenge to find a good trade-off between reporting somewhat suspicious behavior to the back-end in order to allow for better attack detection and not using too much bandwidth.

\paragraph{Comparison and Reproducibility}
\label{par:comparison_reproducibility}

A final important open issue is that many of the presented articles are difficult to compare, because their detection mechanisms rely on different attacker models.
In this article, we attempted to describe the differences in attacker model by illustrating vulnerabilities in some mechanisms, but to truly compare schemes, detection should be performed in the same scenario against the same attacks.
The simplest way to do this would be to use a common dataset, shared between different authors, and this is the approach suggested in~\cite{Rens:VeReMi}.
However, as noted in that article, it would be erroneous to conclude from a high detection rate against this dataset that all attacks can be detected.
To achieve good results, a mutual development of novel attacks and detection mechanisms is necessary.
This in turn requires that attack and detection code be published alongside the articles, such that authors can compare results.
Unfortunately, this practice is not yet commonplace in our field, and thus this presents an important open challenge.

\subsection{Applications beyond cITS}
\label{sub:applications_beyond_cits}

Now that we have studied and classified the state of the art misbehavior detection mechanisms for \acp{cITS}, we address the possible application of the ideas of these mechanisms to other \acf{CPS}.
Although many possible choices exist, we have chosen two domains that seem most appropriate: \acp{WSN}, due to their ephemeral and ad-hoc nature, and \acp{ICS}, due to the potential for (especially) plausibility-based misbehavior detection mechanisms that we have identified.

\subsubsection{Wireless Sensor Networks}
Similar to \acp{cITS}, \acp{WSN} are a type of network that is built up out of many different nodes with significant geographic distribution and connected through a wireless medium.
Contrary to \acp{cITS}, mobility is limited in a \ac{WSN}, as sensors are relatively stationary entities.
The main purpose of \acp{WSN} is to collect information in an efficient and cost-effective manner, using cheap throw-away devices that have a battery and a wireless transmitter.
For example, \acp{WSN} are used to monitor the temperature of the great barrier reef~\cite{Huddlestone-Holmes2007:WSN-GBR}.
For the lifetime of these networks, it is extremely important to monitor resource usage, particularly when transmitting messages, because transmission of messages is by far the most costly operation in terms of battery power.
There are usually also one or more base stations involved in a \ac{WSN}, to provide access to the sensor functionality and to perform more expensive operations~\cite{Zhou2008-WSNSurvey}.

There is already existing work that extensively surveys attacks and defensive techniques for \acp{WSN}, which we briefly discuss below.
We discuss why these approaches cannot be generalized to \acp{CPS}, while the mechanisms we have reviewed show significant potential to be generalized and applied to \acp{WSN}.

Zhou et al.~\cite{Zhou2008-WSNSurvey} describe the \ac{WSN} setting primarily as defined by resource-constrained sensors that perform both sensing and processing, while having a central authority that establishes the network.
The network is mostly static and includes one or more base stations controlled by the authority.
The main security challenges identified by the authors are the public nature of both the wireless medium and the protocols, the lack of (physical) surveillance of nodes and finally the extreme resource constraints.
They also identify node compromise as one of the most challenging attacks on \ac{WSN}, which we have similarly identified as applying to general \acp{CPS}.
Contrary to the \ac{cITS} use case, \acp{WSN} typically also include confidentiality requirements.
The authors then provide a thorough analysis of key management protocols, authentication and integrity, secure routing, intrusion detection and secure applications.
Out of these security mechanisms, we focus on those relating to our discussions regarding reactive security: integrity and intrusion detection, which are the main issues that we see recurring in \acp{CPS}.

As we have emphasized, protecting the message in transit is not sufficient when insider attackers need to be considered.
The authors of~\cite{Zhou2008-WSNSurvey} address the issue of node compromise in their section on intrusion detection.
Although node compromise implies an outside attacker, the effects on the network are essentially the same; an attacker can transmit validly signed message (i.e., this enables misbehavior, which is what we aim to detect).
The discussed misbehavior detection mechanisms raised in this work include a discussion of location-based keys, signal analysis as discussed in Section~\ref{sub:plausibility} and countermeasures to routing misbehavior as discussed in Section~\ref{sub:behavioral_mechanisms}.
Finally, the authors have discussed denial of service and jamming a type of attack that is extremely hard to detect, and secure base stations as another issue that is fairly specific to \acp{WSN}.
The authors note that it is possible that base stations become compromised, even though they are generally assumed to be secure, followed by a discussion of preventive mechanisms for base stations.
Similar assumptions exist in \acp{cITS} for \acfp{RSU}, but the detection of malicious \acp{RSU} is generally easier because they are not essential to a \ac{cITS} the way they are for \acp{WSN}.
Mechanisms employed for \acp{cITS} could be used to address this challenge in exchange for higher computational overhead.

Finally, as their section on open issues describes~\cite{Zhou2008-WSNSurvey}, detecting intruders remains a difficult task, and many mechanisms only focus on a particular attack.
We claim that our survey provides further insight into what mechanisms exist, and we include several mechanisms that describe combinations of existing schemes.
For future work, we believe a primary goal of the \ac{CPS} research community should be to develop a more systematic approach to this problem.
This approach can then be aided by our survey, which can be used to select different types of mechanisms to gain an overall insight of potential attackers in the network.
This includes the transfer of mechanisms described here into the area of \acp{WSN}, in particular the recurring patterns we have identified in the previous section.
For example, physical models that describe the behavior of a complex system monitored by a \ac{WSN} can be used to verify the messages transmitted by individual nodes at a base station, or at an aggregating node.

\subsubsection{Unmanned Areal Vehicles}

Recent advances in the development and wide-spread adoption of unmanned areal vehicles (UAVs) have led to the idea that it may be useful to add communication capabilities to these devices.
In some sense, networked UAVs are thus quite similar to VANETs: both are types of ad-hoc networks with mobile nodes and a decentralized infrastructure.
In contrast to cITS, where the existence of periodic availability of back-end communication and relatively high-power devices, UAVs put much stricter requirements on the decentralized nature of communication.
Additionally, the application of these networks is almost exclusively routing~\cite{UAV-Survey}, and thus the associated attacks fall outside the scope of this survey.
Future applications of UAVs that also include application-specific information (e.g., relative positions and speeds), which are subject to similar attacks addressed by most detectors described in this article, as addressed by some recent work~\cite{Schaefer-WiSec2016}.
Similarly, applications discussed for WSNs discussed above apply equally to UAVs designed for collaborative sensing applications, as such networks are simply a highly dynamic variant of WSNs.
Some dedicated security mechanisms have been proposed specifically for these networks~\cite{martini_distributed_2015,sedjelmaci_hierarchical_2017}; however, adopting these works to the specifics of cITS is likely non-trivial, because these authors specify different assumptions.
For example, these assumptions include a much lower density, no privacy requirements and a central control system that collects the data (the ground station in~\cite{sedjelmaci_hierarchical_2017}).
For a detailed review of security and privacy requirements for these devices, we refer interested readers to an existing survey on this topic~\cite{Altawy-sps-drones}.

\subsubsection{Industrial Control Systems} 
Unlike \acp{cITS}, \acp{ICS} are a type of \acp{CPS} that has historically grown out of the field of control systems, and the focus of these systems has primarily been safety, rather than security~\cite{kargl_insights_2014}.
Specifically, \acp{ICS} must avoid individual but random failures from escalating and causing a \emph{cascading failure}, that is, causing a shutdown of components that are not directly related to the original failure.
To this end, much research has been conducted, but recent developments regarding attacks on these systems, such as those by Stuxnet, that these systems are not built to resist targeted attacks.
In the past, \acp{ICS} were internal networks, which were completely separated from the Internet by a so-called ``air-gap'', meaning no direct connectivity is possible.
However, the necessity of patching these systems and the use of re-writable media like USB-devices has led many security researchers to conclude that a complete ``air-gap'' is not a feasible solution.
In addition, secure connectivity has already provided organizations with significant cost savings and ease of management for these systems.
Therefore, we study how misbehavior detection can be used to improve security in these systems, in particular against an insider attack that attempts to cause damage by disrupting or sabotaging the industrial process controlled by the \ac{ICS}.
Again, the recurring patterns we have identified for \acp{cITS} may provide significant insights for new developments in \ac{ICS}, for example through the application of physical models to detect anomalies with high accuracy.

\section{Conclusion}
\label{sec:conclusion}

In this survey, we have provided an overview of different approaches to misbehavior detection in \acs{cITS}.
\ac{cITS} are a promising type of networked system that connect vehicles, road-side units and back-end systems in order to achieve safer, more efficient and more comfortable travel on roads.
\ac{cITS} are an instance of \ac{CPS}, a type of system where interaction between the physical world and the cyber world is a central aspect.
\ac{CPS} have several unique challenges, including the critical usage scenarios, strong resource and cost constraints, and high scalability requirements.
These challenges lead to new and strong security requirements, which lead to the development of misbehavior detection as a second, reactive layer of security on top of the first, proactive layer.
In the case of \ac{cITS}, this proactive security layer is the \acs{PKI}, which enables the exclusion of attackers that do not possess key material.
However, in highly constrained environments like \ac{CPS}, some network nodes may be compromised to obtain key material; in particular, in \ac{cITS}, an attacker may obtain legitimate key material by manipulating her own vehicle to send arbitrary messages.
Therefore, reactive security in the form of misbehavior detection is a necessary tool to provide security in \ac{cITS}.
Although many of these schemes were originally designed for \acp{VANET}, our discussion is designed to present them in the context of upcoming \acp{cITS}, where autonomous vehicles communicate.
We have surveyed such misbehavior detection mechanisms in detail, and discussed their generalizability towards other \ac{CPS}.

To this end, we have first defined the system model for \ac{cITS} (Section~\ref{sec:system_model}) and the concept of misbehavior in these systems (Section~\ref{sec:misbehavior}).
We then presented a classification of different misbehavior detection mechanisms that have been developed in the literature over the last decade, designed specifically for \ac{cITS}.
The classification consists of node-centric mechanisms, which use properties of a sender to detect malicious messages, and data-centric mechanisms, which primarily analyze the semantics of received messages.
Node-centric mechanisms can be sub-divided in two further classes; behavioral mechanisms, where the receiver analyzes aspects like message frequency and conformance to standards, and trust-based mechanisms, which allow nodes to express trust in messages or senders directly.
Data-centric mechanism can be divided in plausibility mechanisms, analyzing the semantics of a series of packets from an individual sender, and consistency mechanisms, which analyze the consistency of messages received from multiple different senders.
The classification has been used to discuss a large number of different misbehavior detection mechanisms developed in the literature, the results of which are summarized in Tables~\ref{tab:mechAssess1}, \ref{tab:mechAssess2} and \ref{tab:mechAssess3}.

Although we have limited the survey of mechanisms to those designed for \ac{cITS}, we believe that this survey offers a contribution to the wider field of \ac{CPS}, as the novel aspects of mechanisms developed for \ac{cITS} can be used to improve the security of other \ac{CPS}.
To take a first step towards this goal, we have analyzed the common aspects in Section~\ref{sec:solved_and_open_challenges} and discussed their application to other systems from the \ac{CPS}-domain.
We envision two broad paths for future work; for security in \ac{cITS}, taking further advantage of the highly orthogonal nature of the four different classes of misbehavior detection we have presented is of vital importance, while the generalization of misbehavior detection mechanisms for \ac{cITS} to general \ac{CPS} will allow the field to advance without sacrificing security.
This survey has provided a wide scope of different detection approaches, which provides a conclusive overview of and introduction to the field for industry, developers, standardization agencies, and aspiring researchers in this field.

\section*{Acknowledgements}

We thank Anke Jentzsch and Hendrik Decke from Volkswagen AG, Germany, for their cooperation, feedback and fruitful discussions concerning this survey.
This work was supported in part by the Baden-W\"{u}rttemberg Stiftung gGmbH Stuttgart as part of the project IKT-05 AutoDetect of its IT security research programme.

\bibliographystyle{IEEEtran}
\bibliography{IEEEabrv,references}

%

\newpage

\begin{IEEEbiography}[{\includegraphics[width=1in,height=1.25in,clip,keepaspectratio]{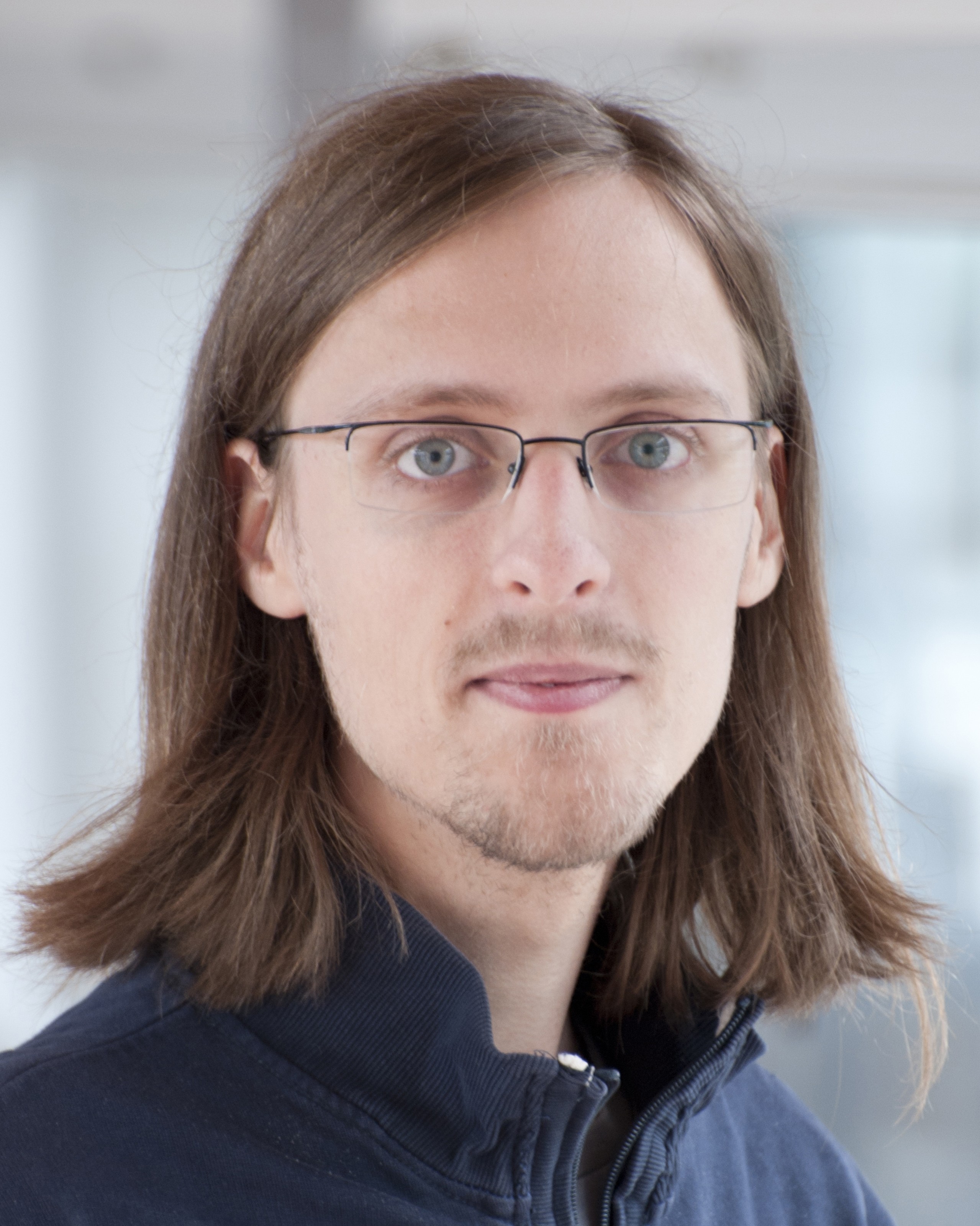}}]{Rens Wouter van der Heijden}
  Rens Wouter van der Heijden is a PhD student with the Institute of Distributed Systems at Ulm University. He earned is Masters' degree at Twente University in 2012, specializing in computer security. His research focuses on misbehavior and intrusion detection in cooperative intelligent transport systems, with additional interests in information fusion, subjective logic, and network security.
\end{IEEEbiography}

\vspace{-0.8em}

\begin{IEEEbiography}[{\includegraphics[width=1in,height=1.25in,clip,keepaspectratio]{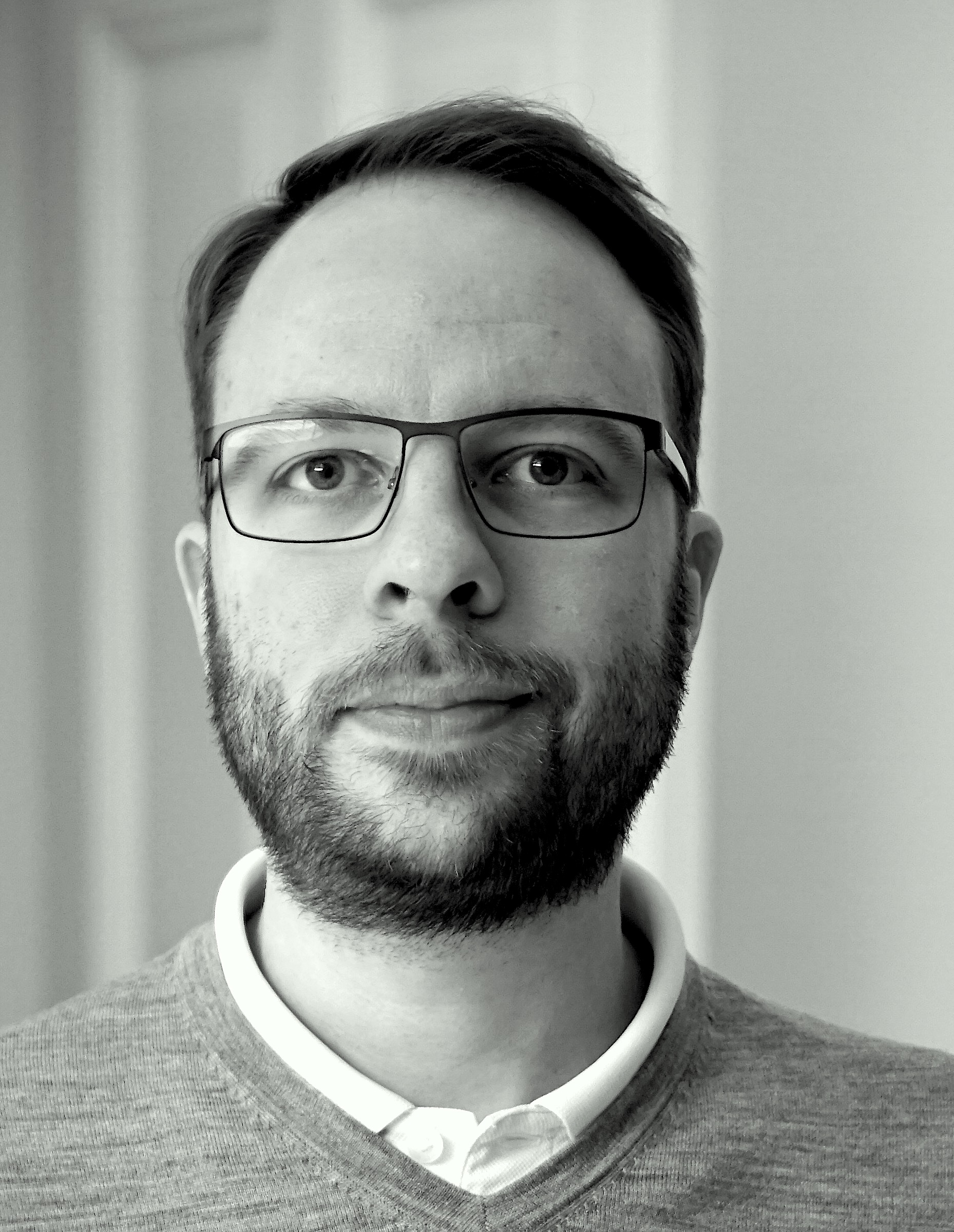}}]{Stefan Dietzel}
Stefan Dietzel is a postdoctoral researcher with the chair of computer engineering at Humboldt University, Berlin, Germany. Stefan holds a doctoral degree in computer science from the University of Twente, The Netherlands, and a Diplom degree in computer science from Ulm University, Germany. His research interests include efficient and resilient wireless ad-hoc communication systems for industrial environments, as well as vehicle-to-vehicle communication.
\end{IEEEbiography}

\vspace{-0.8em}

\begin{IEEEbiography}[{\includegraphics[width=1in,height=1.25in,clip,keepaspectratio]{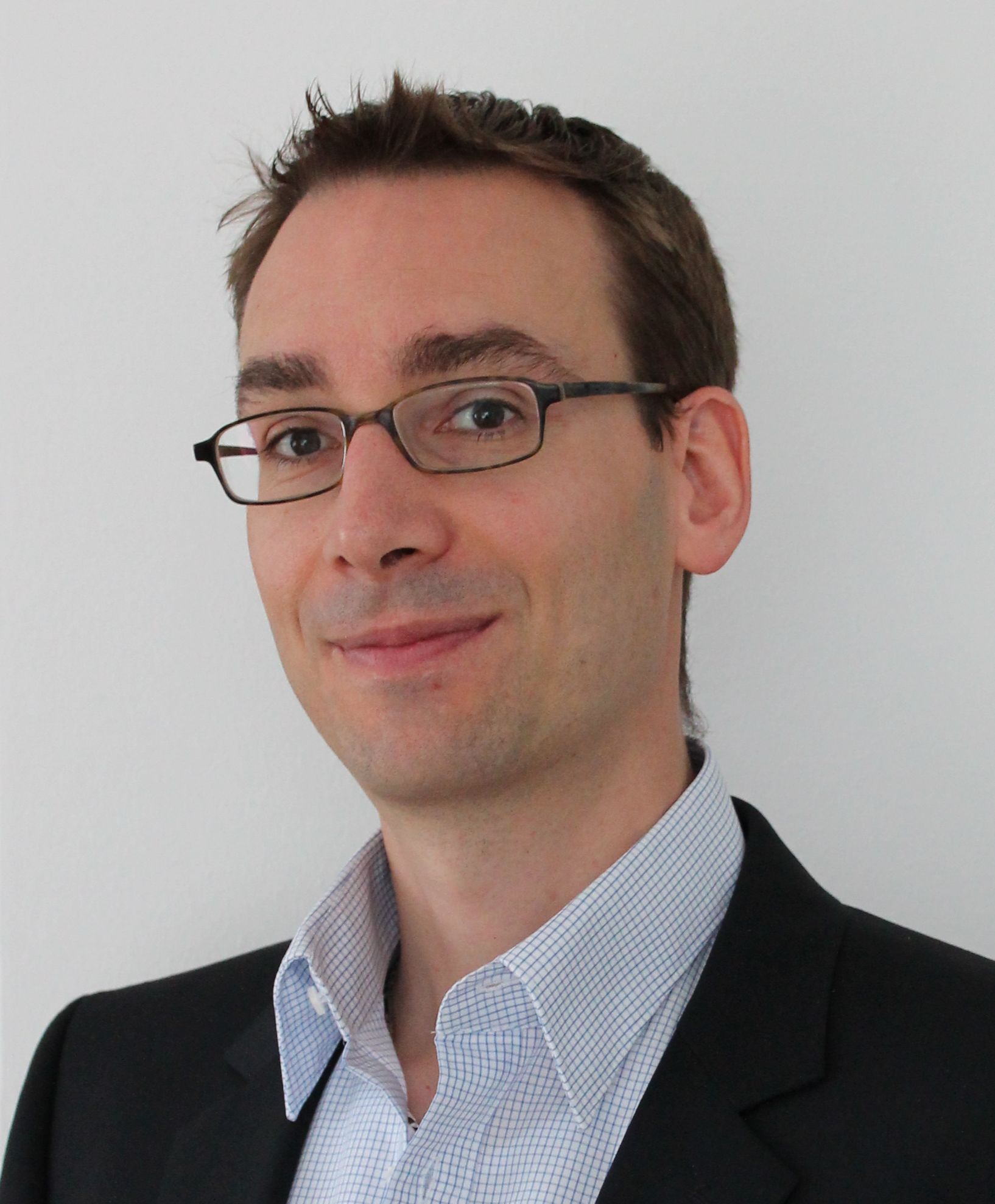}}]{Tim Leinmüller}
  Tim Leinmüller is currently the technical manager responsible for DENSO’s European Connected Automated Driving (CAD) research and standardization activities. He serves as official contact and is responsible for DENSO’s involvement in related industry groups and standardization activities, such as the 5G Automotive Association (5GAA), the Car2Car Communication Consortium (C2C-CC), and ETSI. Tim's research interests and activities include any aspects of connected vehicle technology, as well as automotive (cyber)security.
\end{IEEEbiography}

\vspace{-0.8em}

\begin{IEEEbiography}[{\includegraphics[width=1in,height=1.25in,clip,keepaspectratio]{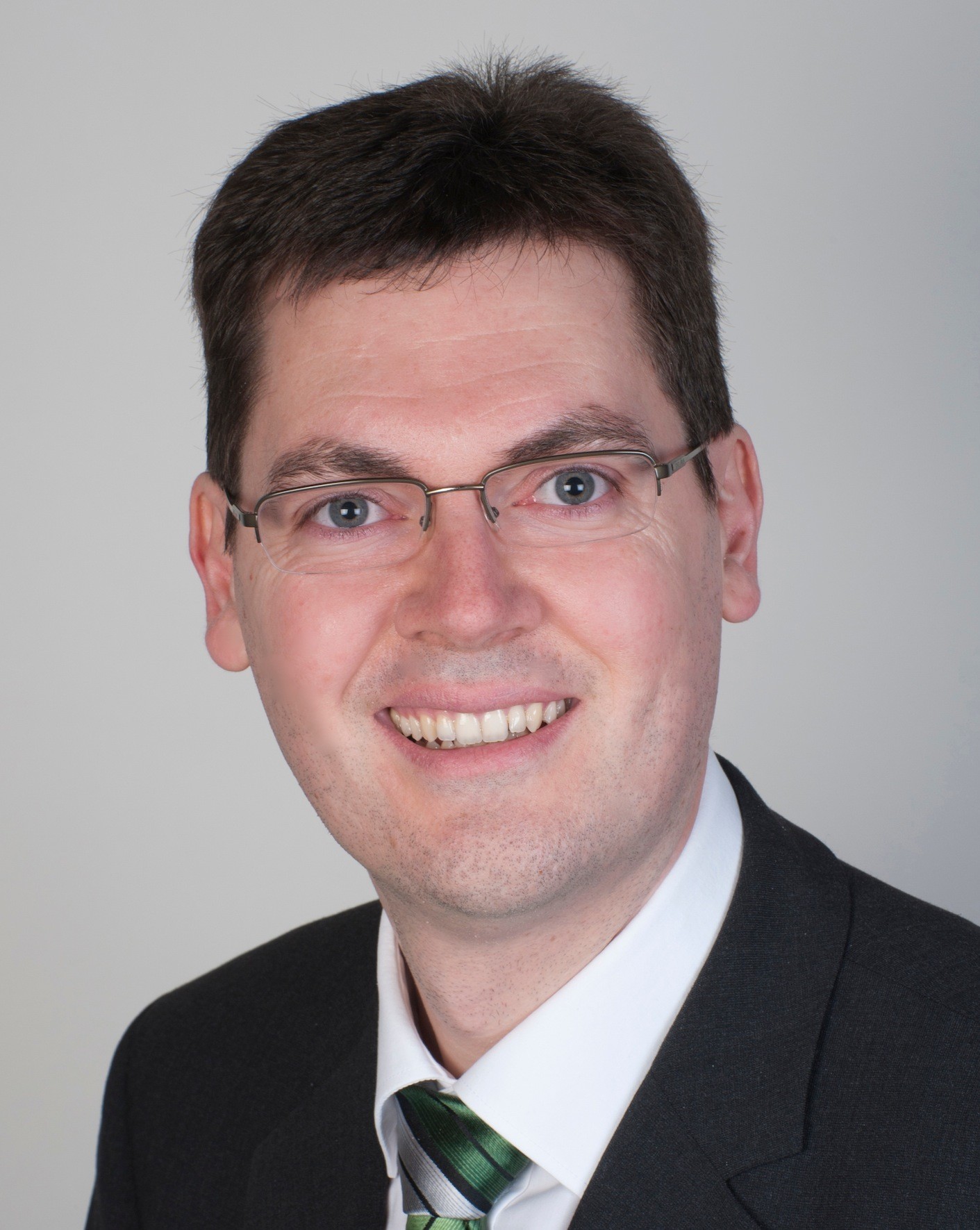}}]{Frank Kargl}
  Frank Kargl holds the chair of Distributed Systems at Ulm University. His research interests include network security and privacy with a special focus on cooperative intelligent transportation systems and automotive systems. Other research includes privacy-enhancing technologies and distributed computing frameworks. In these areas, he co-authored more than 200 peer-reviewed publications. Frank is a member of ACM, IEEE and German GI.
\end{IEEEbiography}
\vfill

\end{document}